\documentclass[mathleft
]{an}
\usepackage{graphicx}
\usepackage{amsmath}
\usepackage{times}
\begin{document}

\Pagespan{1}{}
\Yearpublication{2011}%
\Yearsubmission{2011}%
\Month{1}%
\Volume{1}%
\Issue{1}%

\title{Analytic and numerical models of the 3D multipolar magnetospheres of pre-main sequence stars}

\author{Scott G. Gregory\inst{1}\fnmsep\thanks{Corresponding author:
  \email{sgregory@caltech.edu}\newline}
\and  Jean-Fran{\c c}ois Donati\inst{2}
}
\titlerunning{Multipolar magnetospheres}
\authorrunning{S. G. Gregory \& J.-F. Donati}
\institute{
California Institute of Technology, MC 249-17, 
Pasadena, 91101 CA, U. S. A.
\and 
IRAP-UMR 5277, CNRS $\&$ Univ. de Toulouse, 14 Av. E. Belin, F-31400 Toulouse, France}

\received{20 September 2011}
\accepted{19 October 2011}
\publonline{later}


\abstract{
Traditionally models of accretion of gas on to T Tauri stars have assumed a dipole stellar magnetosphere,
partly for simplicity, but also due to the lack of information about their true magnetic field topologies.  
Before and since the first magnetic maps of an accreting T Tauri star were published in 2007 a new generation
of magnetospheric accretion models have been developed that incorporate multipole magnetic fields.  Three-dimensional models of the large-scale
stellar magnetosphere with an observed degree of complexity have been produced via numerical field extrapolation 
from observationally derived T Tauri magnetic maps.  Likewise, analytic and magnetohydrodynamic models with 
multipolar stellar magnetic fields have been produced.  In this conference review article we compare and contrast the 
numerical field extrapolation and analytic approaches, and argue that the large-scale magnetospheres of some (but not all)
accreting T Tauri stars can be well described by tilted dipole plus tilted octupole field components.  We further argue that 
the longitudinal field curve, whether derived from accretion related emission lines, or from photospheric absorption lines, provides 
poor constrains on the large-scale magnetic field topology and that detailed modeling of the rotationally modulated Stokes V signal
is required to recover the true field complexity.  We conclude by examining the advantages, disadvantages and
limitations of both the field extrapolation and analytic approaches, and also those of magnetohydrodynamic models. 
}

\maketitle


\section{Introduction}
The original models of magnetospheric accretion on to a T Tauri star assumed the stellar magnetosphere was 
a dipole (K{\"o}nigl 1991; Collier Cameron $\&$ Campbell 1993; Hartmann et al 1994; Shu et al 1994).  This assumption 
was made partly for simplicity and partly due to a lack of available 
information about the true large-scale magnetic field topology of such stars.  None-the-less dipole magnetospheric
models have proved successful in reproducing many observational signatures, for example, the profiles
and absorption components of accretion related emission lines (e.g. Hartmann et al 1994; Kurosawa et al 2006).  Since the middle of the 1990s
several two dimensional magnetohydrodynamic (MHD) models have been presented (e.g. Goodson et al 1997; Miller $\&$ Stone 1997; 
Romanova et al 2002; K{\"u}ker et al 2003; von Rekowski $\&$ Piskunov 2006; Bessolaz et al 2008; Zanni $\&$ Ferreira 2009),
as well as some three dimensional (3D) simulations with tilted dipole magnetospheres (Romanova et al 2003, 2004; Orlando et al 2011) - see 
Gregory et al (2010) for a review.

Dipolar accretion models are not without their critiques (e.g. Safier 1998), and dropping the dipole assumption can significantly affect
the structure of accretion columns within the magnetosphere (Gregory et al 2006a), the locations of hot spots at the base of accretion funnels 
(Gregory et al 2005; Mohanty $\&$ Shu 2008), and the density of the gas in the flow (Gregory et al 2007).  High order field components may even 
play a dominate role in the physics of the gas inflow as the accretion columns approach the star (Adams $\&$ Gregory 2011).  

Through the analysis of Zeeman broadened lines in intensity spectra T Tauri stars have been found to host surface 
averaged magnetic fields of order a kilo-Gauss (see Johns-Krull 2007 and references there-in).
Provided that the surface fields are associated with a stellar magnetosphere that is sufficiently globally ordered then such
strong fields are able to disrupt the circumstellar
disk at a distance of a few stellar radii, as required by the magnetospheric accretion scenario (K{\"o}nigl 1991).  There 
appears to be little variation in the average surface field between sources, and between accreting and non-accreting stars,
although intriguingly the average (unsigned) magnetic flux appears to be less, on average, for stars in older star forming 
regions (Yang $\&$ Johns-Krull 2011).  

Spectropolarimetry provides a method of probing the magnetic field topology.  By measuring the circular polarization signal 
in magnetically sensitive photospheric absorption lines, and in accretion related emission lines, information
can be obtained about the surface fields of T Tauri stars.  Zero net polarization signals have often been measured in photospheric absorption lines 
(e.g. Valenti $\&$ Johns-Krull 2004).  
As such lines form essentially uniformly across the entire visible stellar surface, this suggests that T Tauri stars host complex surface fields with 
roughly equal amounts of positive and negative field regions.  However, often accretion related emission lines, such as HeI 5876{\AA }, 
show strong levels of circular polarization (e.g. Johns-Krull et al 1999), with the signal modulated on a timescale equal to the photometrically
determined rotation period of the star (Valenti $\&$ Johns-Krull 2004).  This suggests that for many T Tauri stars magnetospheric accretion occurs predominantly 
into strong single polarity spots on the surface the star.  However, such studies which are based on measuring the phase modulated
stellar disk averaged longitudinal field component, cannot constrain the field topology of the star beyond the interpretation that T Tauri stars
have complex surface fields, and somewhat simpler large scale fields.     

In 2007 the first magnetic maps of the large-scale field of an accreting T Tauri star were published (Donati et al 2007).  
At the time of writing magnetic maps have now been produced for 10 accreting T Tauri stars namely V2129~Oph, BP~Tau, V2247~Oph, AA~Tau,
TW~Hya, both stars of the close binary V4046~Sgr, CR~Cha, CV~Cha and MT~Ori (Donati et al 2007, 2008b, 2010a,b, 2011a,b,c; 
Hussain et al 2009; Skelly et al 2011).  They are produced by monitoring the rotational modulation of the circular polarization signal 
over a complete stellar rotation cycle, and in practise several, detected in both photospheric absorption and accretion related emission lines.
As the polarization signal in individual photospheric absorption lines is vanishingly small cross-correlation techniques (such as Least-Squares 
Deconvolution (LSD); Donati et al 1997) are employed in order to extract information from as many of the spectral lines as possible.
The complete process, known as Zeeman-Doppler imaging, see Donati (2001) for a review and Donati et al (2010b) for its specific application to 
accreting T Tauri stars, simultaneously reconstructs maps of the radial, azimuthal and meridional field components at the stellar surface, 
as well as a brightness map (the surface distribution of cool spots) and a map of excess emission in accretion related lines (the surface 
distribution of accretion hot spots).  Due to the flux cancellation effect that affects all polarization measurements, the magnetic maps only 
contain information about the large scale field topology.  Once the magnetic maps have been derived the field components can be written in 
terms of spherical harmonics (e.g. Donati et al 2006b).  From the maps the coefficients of the spherical harmonic decomposition can be calculated, 
and from these the polar strength of the each of the multipole moments along with their tilts relative to the stellar rotation axis, and the 
phase which they are tilted towards, can be determined.  A description of the observational process involved in constructing 
stellar magnetic maps via Zeeman-Doppler imaging, as well as the assumptions and limitations, can be found elsewhere in this 
volume in the paper of G. A. J. Hussain, to which readers are referred for further details (see also Donati $\&$ Landstreet 2009).    

Most of the published magnetic maps have been obtained as part of the Magnetic Protostars $\&$ Planets (MaPP) program.
This ongoing large program consists of 690 hours of observing with the ESPaDOnS (Echelle SpectroPolarimetric Device for the Observation of 
Stars; Donati et al 2006a) instrument at the Canada-France-Hawai'i Telescope (CFHT) between 2008 and 2012, 
1130 hours with the NARVAL spectropolarimeter (Auri{\`e}re 2003) at the T{\'e}lescope Bernard Lyot (TBL) in the Pyren{\'e}es, and additional ground and space based 
observations (e.g. XMM-Newton, Chandra and HST) for select sources.  

In this paper we discuss the results from the ongoing MaPP program and present magnetic maps for some accreting T Tauri stars in 
section \ref{sec_maps}.  We demonstrate that the large scale field of some (but not all) T Tauri stars are well described by fields consisting of a tilted 
dipole plus a tilted octupole component.  We review the current 3D field extrapolation and analytic models of their 
magnetospheres in section \ref{multipole_models}, and argue that use of the stellar disk averaged longitudinal field component alone provides poor constrains on the 
magnetic field topology.  We conclude in section \ref{conclu} by comparing both approaches to one another, and also to
magnetohydrodynamic models, discussing the advantages, disadvantages and limitations of each.


\section{Surface magnetic maps}\label{sec_maps}

\begin{figure*}
\centering
\includegraphics[width=120mm]{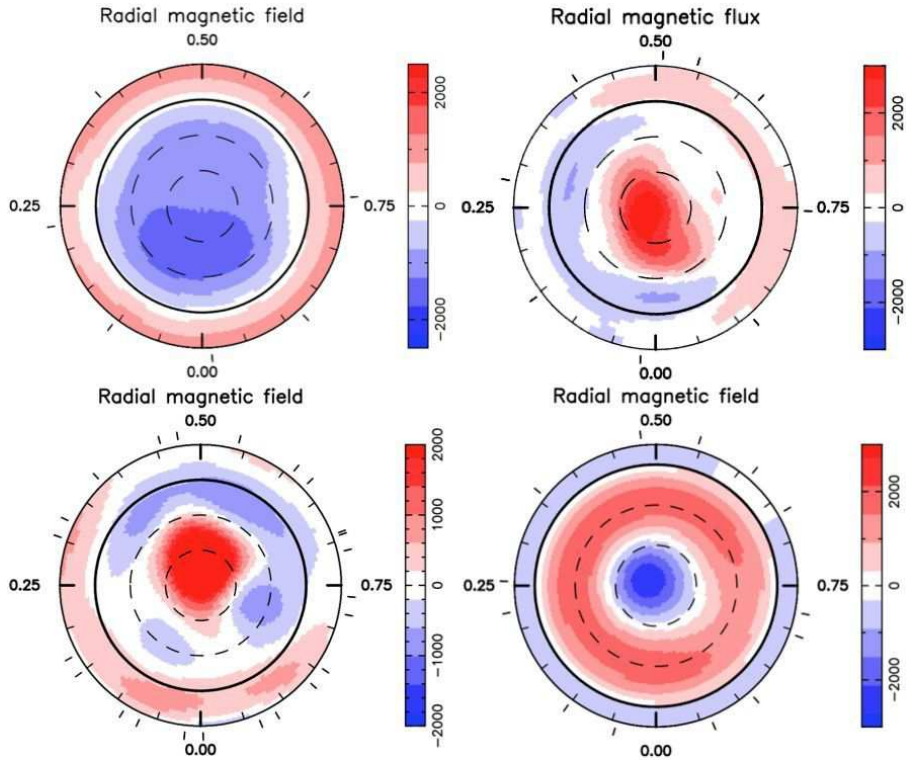}
\caption{Magnetic maps showing only the radial component of the field of four accreting T Tauri stars derived from Zeeman-Doppler imaging.
              The large-scale fields of AA~Tau in January 2009 (upper left; Donati et al 2010b), BP~Tau in February 2006 (upper right; Donati et al 2008), 
              V2129~Oph in July 2009 (lower left; Donati et al 2011a) and TW~Hya in March 2010 (lower right; Donati et al 2011b) are multipolar, but well 
              described as dipole-octupole composite fields with different polar strengths
              and moment tilts for each star (see Table \ref{table}).  The maps are shown in flattened polar projection with positive/negative 
              field regions in red/blue and 
              labeled in Gauss.  The stellar rotation pole is in the middle of each map, the dotted lines are lines of constant latitude separated by $30^\circ$
              with the equator as the bold circle.  The numbers and tick marks around the circumference denote phase and phases of observation 
              respectively.  Other T Tauri stars possess more complex large-scale magnetic fields, see Figure \ref{complexsurfacemaps}.}
\label{surfacemaps}
\end{figure*}

\begin{table*}
 \centering
\caption{Polar field strengths of the dipole and octupole components, their tilts relative to the visible stellar rotation pole ($\beta$), and 
the rotation phase that they are tilted towards ($\Phi$) for the stars show in Fig. 
\ref{surfacemaps}.  The values listed for BP Tau were derived using an experimental
version of the magnetic imaging code and have recently been revised from the values shown here 
(see footnote 1 for details).  The fields of TW Hya and V2129 Oph were significantly different at other observing epochs.}
\label{table}
\begin{tabular}{ccccccccc}\hline
Star & Date & $B_{dip}$ & $\beta_{dip}$ & $\Phi_{dip}$ & $B_{oct}$ & $\beta_{oct}$ & $\Phi_{oct}$ & $B_{oct}/B_{dip}$ \\ 
       &           &   (kG)       &                       &                       &  (kG)        &                       &                     &                              \\
\hline
AA Tau       & Jan09 & 1.7   &   170$^\circ$    &     0.65      &    0.5         &      10$^\circ$          &      0.50           &   0.29           \\
BP Tau       & Feb06 & 1.2   &   20$^\circ$       &    0.65       &   1.6          &      10$^\circ$        &      0.15           &   1.33                   \\
V2129 Oph & Jul09 & 1.0   &   15$^\circ$      &     0.40       &    2.2         &      20$^\circ$          &     0.50           &   2.20                   \\
TW Hya      & Mar10 & 0.7  &   5$^\circ$        &     0.25       &   2.8         &      $>$175$^\circ$   &     0.85          &   4.00                  \\
\hline
\end{tabular}
\end{table*}

T Tauri stars are found to host multipolar magnetic fields, but the field topology varies between stars and over time for the same star.
All appear to have dipole components that are strong enough to disrupt their circumstellar disks.
Although their large scale fields are non-dipolar, they are often simpler than the highly complex fields of rapidly rotating
zero-age main sequence stars (e.g. Donati et al 2003).  This result is not a phase coverage effect, as demonstrated by Hussain et al (2009).
Some T Tauri stars stars host simple axisymmetric fields with strong kilo-Gauss dipole components (e.g AA Tau and BP Tau).  Others host dominantly 
axisymmetric fields but field modes of higher order than the dipole are found to be the most dominant (typically, but not exclusively, the octupole component), 
and their dipole components may be strong or weak (e.g. TW Hya, V2129 Oph and MT Ori).  Others host complex magnetic fields that are highly 
non-axisymmetric with very weak dipole components ($<\sim0.1\,{\rm kG}$; e.g. V4046 Sgr AB, CR Cha and CV Cha).  There is strong link with stellar 
internal structure where fully convective stars close to the end of the fully convective phase host simple fields with strong dipole components.  
The dipole component decays rapidly with the development of a radiative core, although the large scale field appears to retain its axisymmetry
until the core occupies a substantial volume of the star.  Empirically once the mass of the core exceeds about 40\% of the stellar mass, the field
becomes highly complex and the dipole component is weak (see Gregory et al 2012 for full details of how T Tauri magnetic topologies vary with 
stellar internal structure).  There is also tentative evidence for a bi-stable dynamo process operating amongst the lowest mass fully convective 
T Tauri stars, where we expect some to host complex fields, and others simple fields with strong dipole components like the more massive 
fully convective pre-main sequence stars.  This behavior mirrors the magnetic topology trends already found for main sequence M-dwarfs on 
either side of the fully convective limit (Donati et al 2008a; Morin et al 2008, 2010, 2011). 

\begin{figure*}
\centering
\includegraphics[width=120mm]{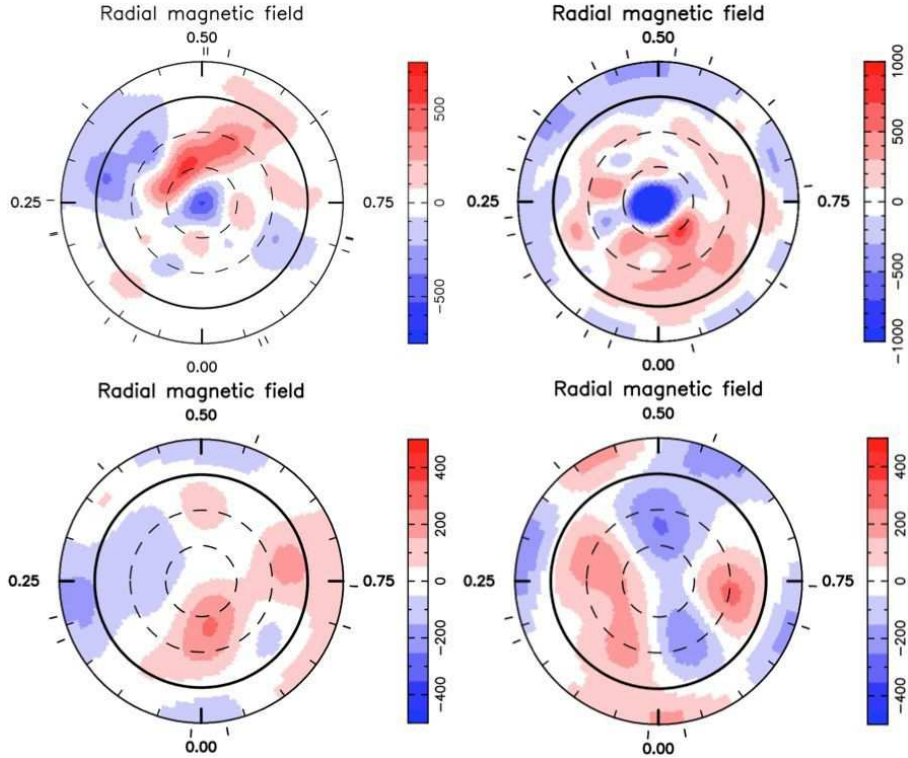}
\caption{As Figure \ref{surfacemaps} but showing magnetic maps of accreting T Tauri stars with large scale fields 
              that are not well described as dipole-octupole composite fields; V2247~Oph in July 2008 (upper right; Donati et al 2010a),
              MT~Ori in December 2008 (upper left; Skelly et al 2011), and both stars of the close binary V4046 Sgr in September 2009 
              (Donati et al 2011c), the primary (lower left)
              and the secondary (lower right).  The field of MT~Ori is similar to that of TW~Hya and V2129~Oph in that its field
              is dominantly axisymmetric but field modes of higher order than the dipole dominate (in this case there are roughly equal contributions from
              the octupole [$\ell=3$], the dotriacontapole [$\ell=5$] and the $\ell=7$ field modes).  The fields of the other
              stars are complex and non-axisymmetric with weak dipole components, as are the fields of CR~Cha and CV~Cha (Hussain et al 2009; 
              maps not shown here).}
\label{complexsurfacemaps}
\end{figure*}

Figure \ref{surfacemaps} shows four examples of magnetic maps of accreting T Tauri stars.  Only the radial field component is shown, and we 
have deliberately selected those stars which have large-scale field topologies that are somewhat simple and well described as dipole-octupole
fields, as these will be useful for our analytic work in section \ref{analytic}.  Other stars have significantly more complex fields, see Figure 
\ref{complexsurfacemaps}.  Maps of the other field components along with brightness and excess accretion related emission 
maps can be found in the appropriate papers.

The four stars shown in Figure \ref{surfacemaps} have dipole and octupole components of differing polar strength, and tilts relative to the 
stellar rotation axis, as summarized in Table \ref{table}.\footnote{The values listed in Table \ref{table} for BP Tau were derived
before the MaPP program began using an experimental version of the magnetic imaging code.  A preliminary reanalysis of this
older BP Tau data using a more mature version of the code (to presented alongside newly obtained data and new magnetic maps) suggests
that $B_{dip}=1.3\,{\rm kG}$, $\beta_{dip}=5^\circ$, $\Phi_{dip}=0.85$, $B_{oct}=1.8\,{\rm kG}$, $\beta_{oct}=5^\circ$ and $\Phi_{oct}=0.15$ 
are more appropriate for this star in February 2006.  As the new data has not yet been published we have used the old values and 
magnetic map for BP Tau in this paper.}  In this work we define the tilt $\beta$ of the dipole or octupole moment to be that of the main
positive pole relative to the visible rotation pole of the star.  Therefore a star which has the main negative pole of the dipole coincident with
the visible rotation pole would have $\beta_{dip}=180^\circ$, and likewise $\beta_{oct}=180^\circ$ would correspond to the main negative pole of the octupole
coinciding with the visible rotation pole.  This is applicable to AA Tau and TW Hya which are close to a configuration with anti-parallel dipole and
octupole component at the epochs listed in Table \ref{table}.  An alternative notation would be to define the tilts of the components according 
to whichever main magnetic pole was in the visible hemisphere of the star, and then list the polar strength of that component as negative. 
For example, using this alternative approach for AA Tau then $B_{dip}=-1.7\,{\rm kG}$, $\beta_{dip}=10^\circ$ and $\Phi_{dip}=0.15$ (since
tilting the negative pole of the dipole by $10^\circ$ towards phase $0.15$ is equivalent to tilting the positive pole 
by $170^\circ$ towards phase $0.15+0.5=0.65$).  Conservative estimates of the uncertainties in the tilts ($\beta$) and the phases of the tilts ($\Phi$)
listed in Table \ref{table} are $\sim10^\circ$ and $0.1$ (especially for moments with small tilts relative to the stellar rotation axis) respectively.

All four stars in Figure \ref{surfacemaps} and Table \ref{table} have been observed at two different 
epochs.  V2129 Oph showed significant changes in the strength of the dipole and octupole field components between observing epochs
with the dipole component increasing by a factor of about three between 2005 and 2011 (from $0.3\,{\rm kG}$ to $0.9\,{\rm kG}$, see Donati et al 2011a).  
Likewise for TW Hya where the dipole component appeared to flip from roughly anti-aligned with the octupole component at one epoch, to being
titled by about $45^\circ$ from an anti-parallel configuration at the other epoch; although as noted in Donati et al (2011b) 
this may be due to the poor phase coverage at one epoch and
requires additional observations to determine if such large scale topology changes are genuine and common for TW Hya.  AA Tau and BP Tau
showed little change in their field topology between observing epochs, save an overall quarter phase shift in the entire map of BP Tau.  This
was most likely due to a small error in the assumed rotation period building up over the roughly 40 stellar rotations between observing
epochs.  Changes in the field topology over time for both stars cannot be ruled out and further observations are required.  
We note that the magnetic maps of BP Tau were constructed and published using an experimental version of the magnetic
imaging code for accreting T Tauri stars.  The Zeeman-Doppler imaging process for such stars involves the use of 
polarization information in accretion related emission lines, and therefore differs from the magnetic imaging process as it applies
to other stars which lack such proxies.


\section{Multipolar magnetospheric models}\label{multipole_models}
In anticipation of the newly available data stream from the ESPaDOnS spectropolarimeter and the observationally
derived magnetic maps of accreting pre-main sequence stars, Gregory et al. (2005) produced the first simulations 
of the magnetospheres of T Tauri stars with complex magnetic fields.  The magnetic maps of young rapidly rotating 
zero-age main sequence stars were used to generate 3D field extrapolations which where then surrounded by accretion disks.
Consequently Jardine et al (2006) and Gregory et al (2006a,b) demonstrated that magnetic fields with an observed degree of complexity
were required to explain many observational results, from the increase in X-ray luminosity with stellar mass,
to the common detection of rotationally modulated X-ray emission (Flaccomio et al 2005).  Subsequently 3D MHD 
simulations of the star-disk interaction with non-dipolar fields were published by Long et al (2007, 2008) 
using dipole-quadrupole 
fields, by Long et al (2009) using dipole-octupole fields, and by Romanova et al (2011) and Long et al (2011) using 
dipole-octupole fields with polar field strengths
and tilts that match those derived from Zeeman-Doppler imaging studies of the accreting T Tauri stars V2129~Oph and 
BP~Tau.  In addition to the field extrapolation models, and the MHD simulations, multipolar magnetic fields can also 
be handled analytically (or at least semi-analytically), which allows for greater clarity of understanding of the physical results 
from numerical models (Gregory et al 2010).  Mohanty $\&$ Shu (2008) have also presented a semi-analytic generalization 
of the Shu X-wind model which drops the dipole magnetosphere assumption.

The spectropolarimetric results, and the derived magnetic maps, provide clear motivation for constructing 3D
models of the non-dipolar magnetospheres of pre-main sequence stars. 
In this section we provide a detailed review of two of the three approaches to handling multipolar stellar magnetic fields, namely field extrapolations
from magnetic maps and analytic models (we compare these to MHD models in section \ref{conclu}). 
Although our attention is focussed on the magnetic fields of accreting pre-main sequence stars
the results that we derive and discuss herein are applicable generally to the magnetosphere of any star or planet.


\subsection{Field extrapolations from magnetic maps}
Stellar magnetic maps derived from Zeeman-Doppler imaging can be used as inputs to 
models of the 3D large-scale magnetosphere.  Numerical field extrapolations
from magnetic maps have been produced for stars of various spectral type and 
evolutionary stage, for example, zero-age main sequence rapid rotators (e.g. Jardine et al 2002); accreting T Tauri stars 
(Jardine et al 2008; Gregory et al 2008); solar-type post T Tauri stars (Dunstone et al 2008; Marsden et al 2011); 
a massive B star (Donati et al 2006b); and for the global solar magnetic field (e.g. Riley et al 2006; Ruan et al 2008).
The potential field source surface method is used to produce the
field extrapolations, a technique originally developed by the solar physics community (Altschuler $\&$ Newkirk 1969; Schatten et al 1969).  
A potential field represents the lowest possible energy state of the stellar magnetic field.  Twisting of the field line by stellar
surface transport effects (surface differential rotation, meridional flow etc) leads to a 
departure from the potential state and allows magnetic energy to be stored in the field which
can eventually be released through magnetic reconnection events that trigger flares. 

The potential field source surface model has two boundary conditions.  The first is that the radial
field component at the stellar surface is the same as determined from the observationally
derived magnetic map of the star whose magnetosphere is being modeled (the magnetic map is thus used as a direct input).
The second boundary condition is that the field is purely radial ($B_\theta=B_\phi=0$) at a spherical
equipotential surface at some height above the star.  This surface, of radius $R_S$, is known as the source
surface and provides a simple method of mimicking the stellar wind plasma distorting and 
pulling open the closed field beyond a certain height above the star.\footnote{Potential field source surface
models have been produced with non-spherical source surfaces (e.g. Schulz et al 1978). In the case of the Sun the source surface
is a prolate spheroid at solar minimum with major axis along the solar rotation axis.  At solar maximum the source
surface is more complex but is approximately spherical once averaged over polar and azimuthal angles (Riley et al 2006).}  The 
source surface thus represents the maximum extent that a closed field line loop can have, and adjustment of
this boundary affects the amount of open and closed flux through the stellar surface reconstructed
in the field extrapolation model.  Smaller (larger) values of $R_S$ yield more (less) open field relative to closed field
(see figure 1 of Gregory 2011). 

In the solar case the source surface radius is constrained by in-situ heliospheric satellite observations to be 
$R_S\sim2.5\,{\rm R}_\odot$.  For stars such in-situ satellite observations are not available and other
indirect methods must be used to constrain the radius of the source surface.  Although various ideas and suggestions
have been put forward, $R_S$ remains a poorly constrained parameter for stars other than the Sun.  However,
once a value for $R_S$ has been chosen, a coronal X-ray emission model can be used to calculate the global
X-ray emission measure for comparison with independently obtained X-ray observations (Jardine et al 2006; Hussain et al 2007).  Adjustments in $R_S$
affect the amount of closed field, and consequently the amount of X-ray bright regions on the star.  However, once
$R_S$ exceeds a few stellar radii the change in the calculated X-ray emission measure
(or equivalently the X-ray luminosity) is small for successively larger $R_S$ values (Jardine et al 2008).  For accreting T Tauri
stars surface hot spots, which form at the shocks at the base of accretion columns, can be used as an additonal
constraint on $R_S$.  Maps of the excess emission in certain accretion related emission lines are constructed as
part of the Zeeman-Doppler imaging process (see Donati et al 2010b).  Current observations suggest that for the majority of accreting
T Tauri stars hot spots are at high latitude towards the visible rotation pole (see also Petrov et al 2011).  
In order to explain how material can accrete from the disk, in the stellar midplane, into high latitude hot spots the source surface 
must be large enough to ensure that closed field can reach towards the polar regions. 

Potential fields are current free and therefore at every point exterior to the star 
$\nabla\times\mathbf{B}=\mathbf{0}$.  Assuming that the field external to the star can be written as the 
gradient of a magnetostatic scalar potential $\Psi$, $\mathbf{B}=-\nabla\Psi$, then this combined 
with Maxwell's equation that the field is solenoidal ($\nabla\cdot\mathbf{B}=0$) means that the 
magnetostatic potential at any point $(r,\theta,\phi)$ exterior to the star can be determined
by solving Laplace's equation $\nabla^2\Psi=0$.  Thus the spherical field components (e.g. 
$B_r=-\partial\Psi/\partial r$ etc) can be calculated at each point exterior to the star, and then
a field line tracing algorithm employed to reconstruct the 3D global magnetic field 
of the star.  Jardine et al (2002) provide further details of the potential
field source surface method as applied to stellar magnetic maps, with Jardine et al (2006) and Gregory et al (2006a, 2008) providing 
details of its application specifically to T Tauri stars.  Potential field models produce unique 3D
field topologies (Aly 1987), once a source surface radius has been selected, however the fields are static in time.  Such models 
cannot be used to model the time evolution of the magnetic field which will evolve due to motion of the foot 
points on the stellar surface (surface transport effects) and by the twisting of the large field due to the interaction with the disk.
We return to these points in section \ref{conclu}.

\begin{figure}
\centering
\includegraphics[width=47mm]{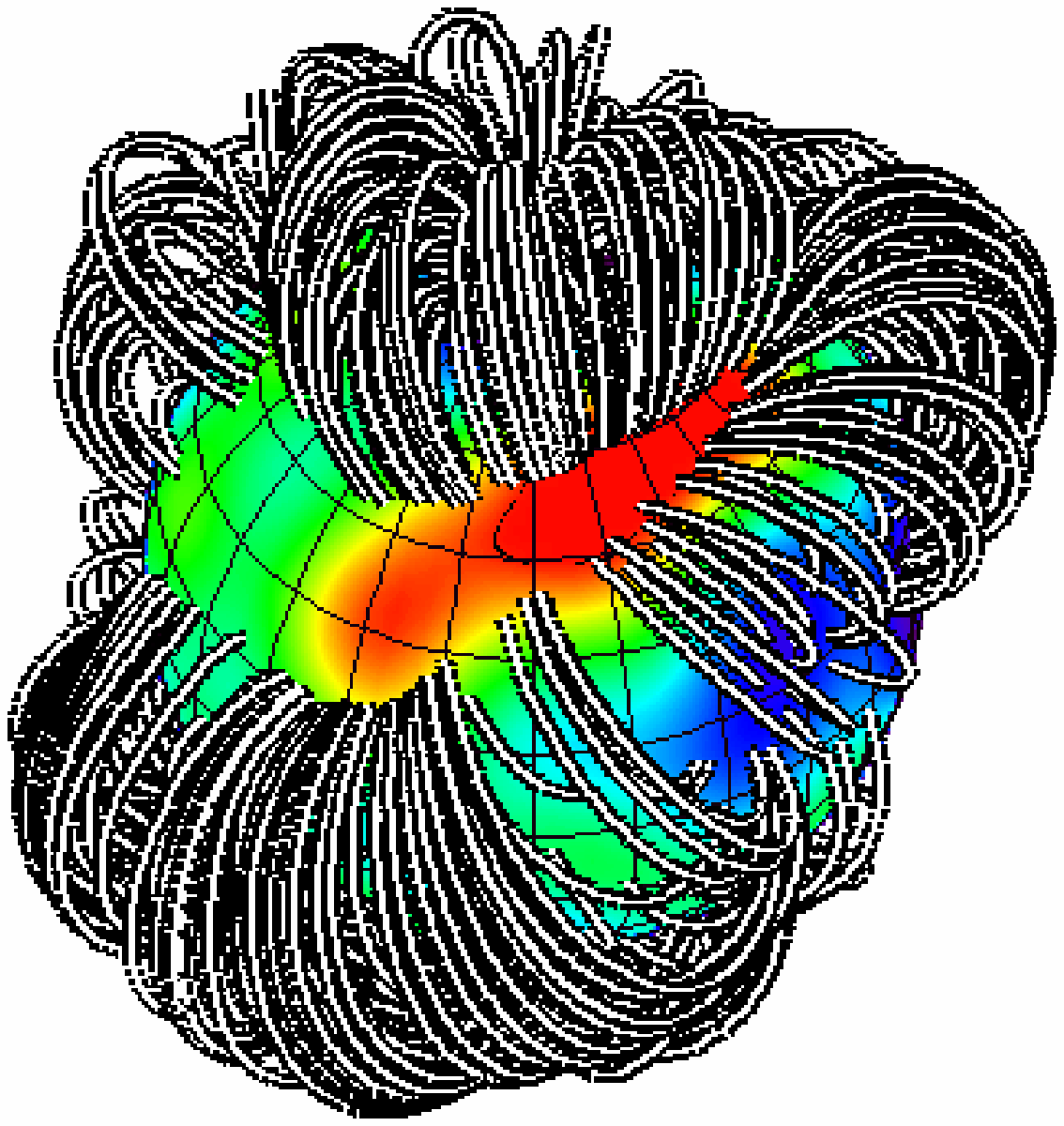}
\includegraphics[width=47mm]{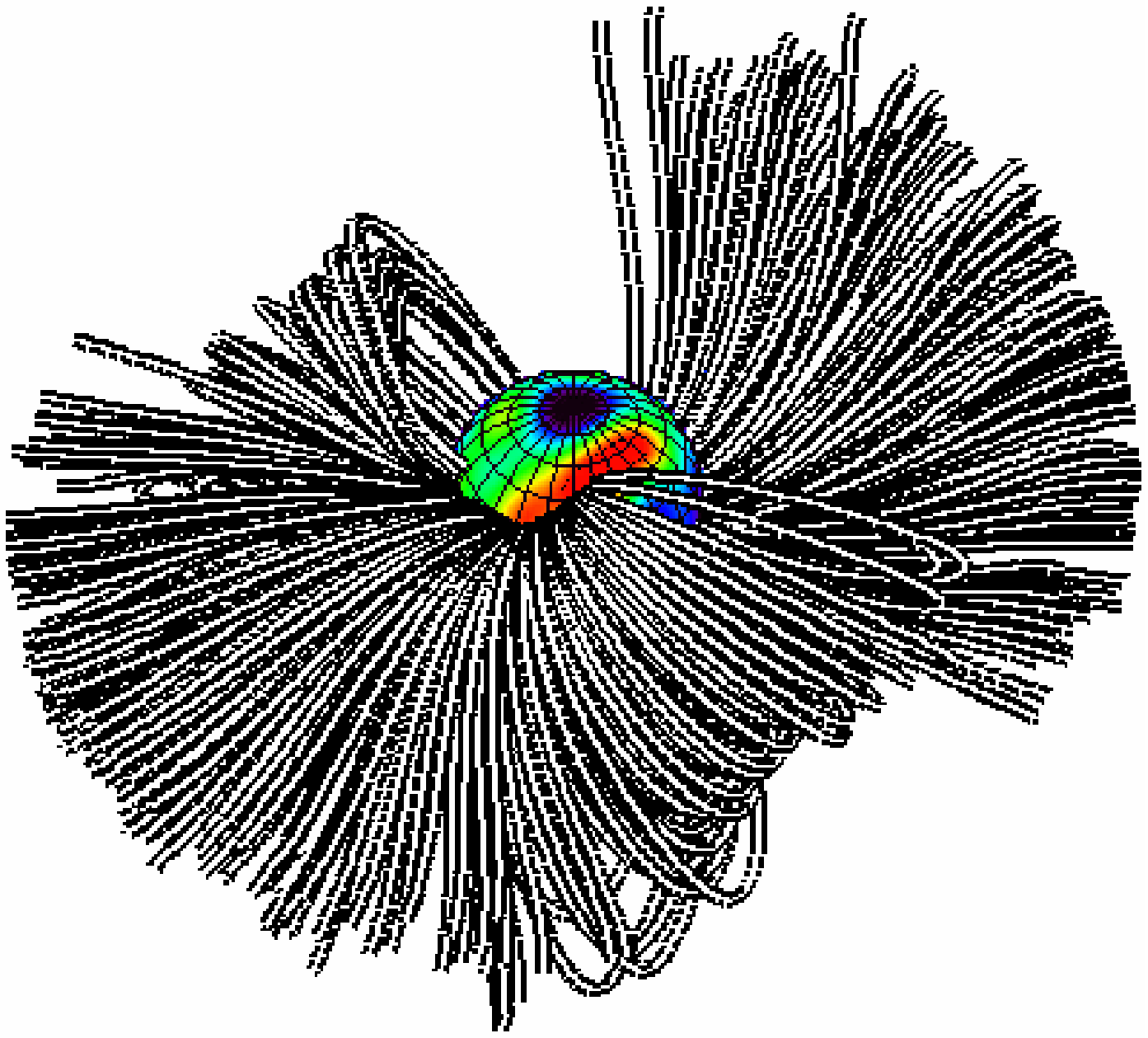}
\includegraphics[width=47mm]{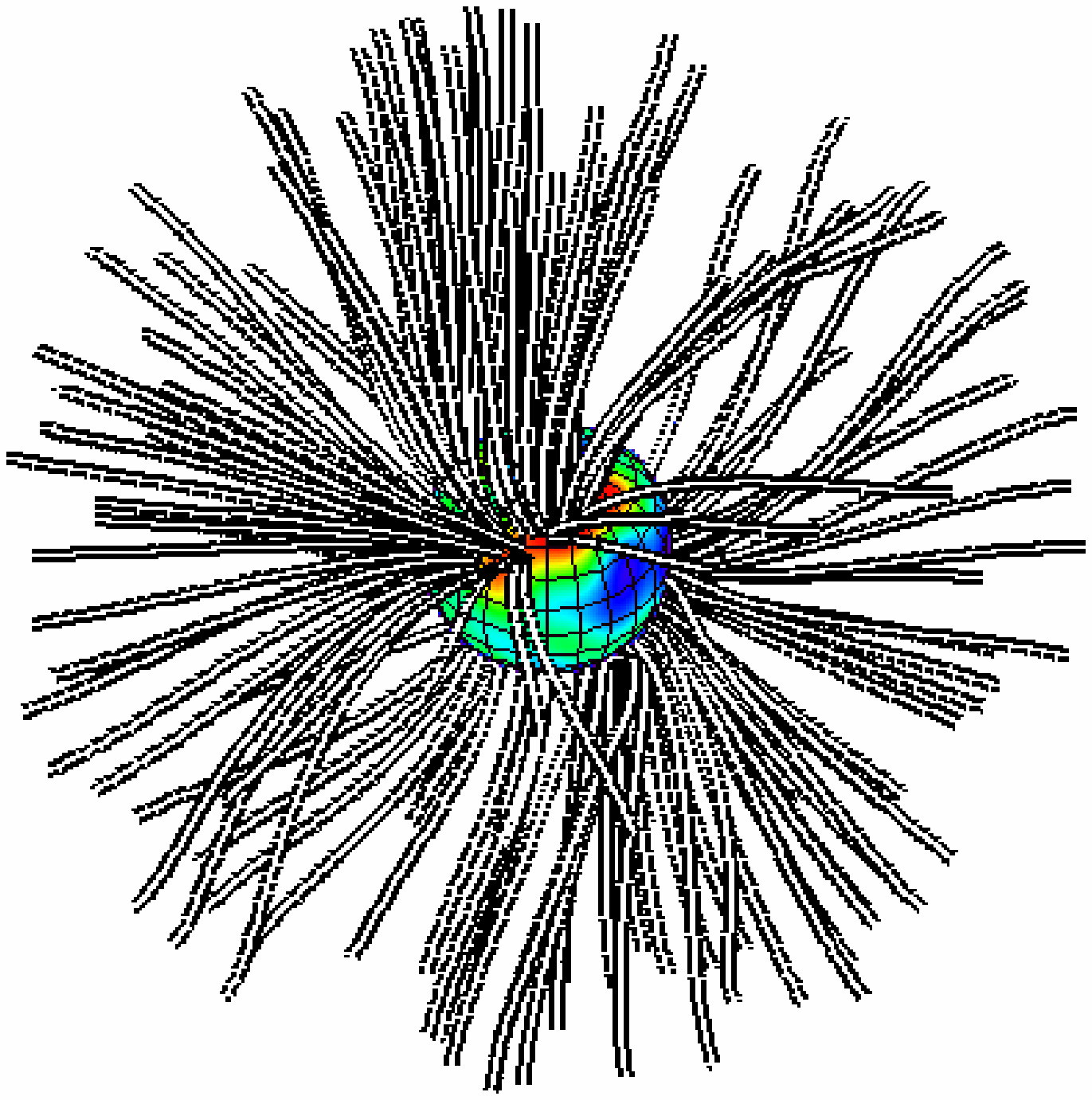}
\includegraphics[width=83mm]{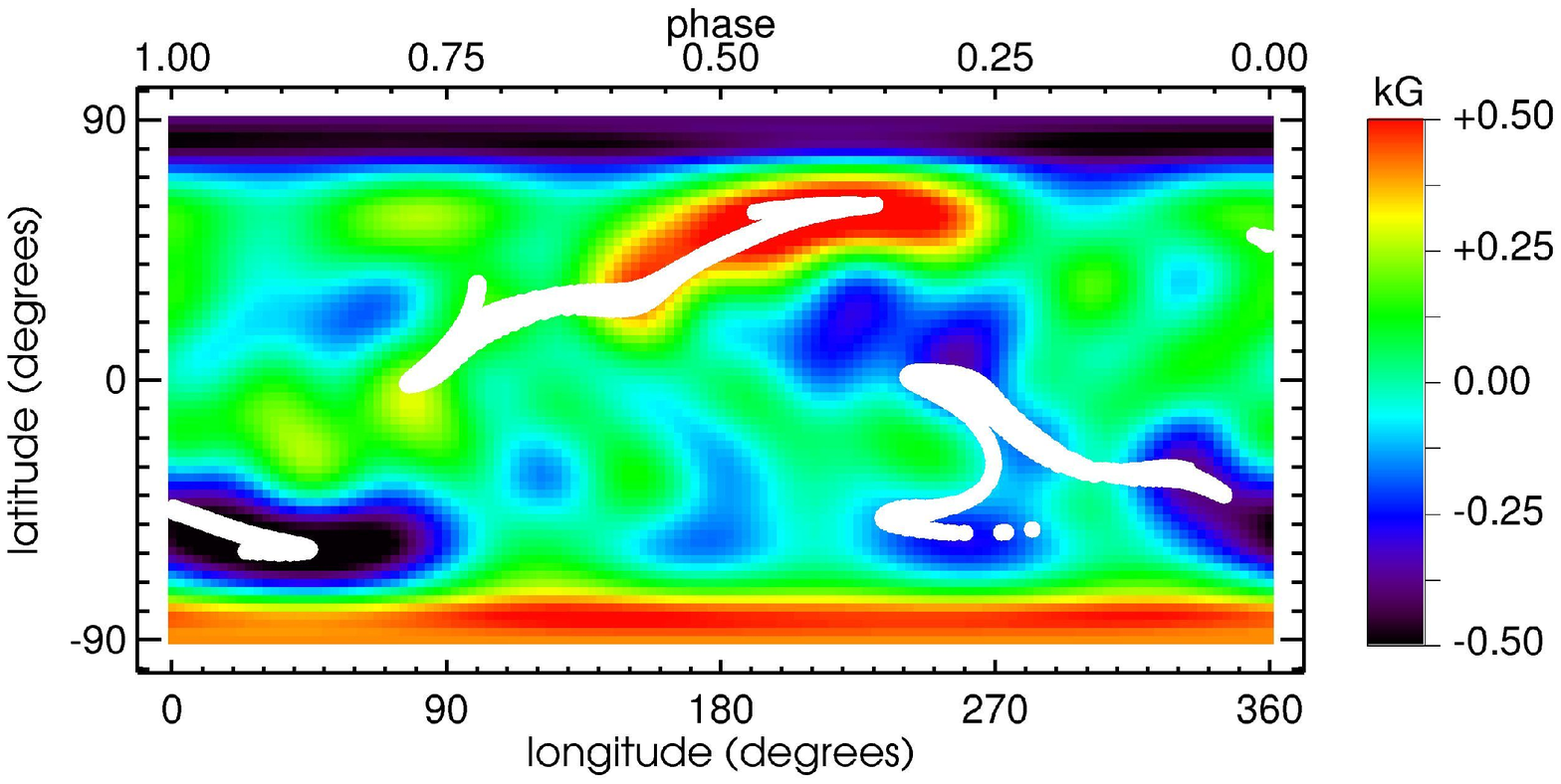}
\caption{Field extrapolations from the magnetic map of V2247 Oph, shown in Figure \ref{complexsurfacemaps} and the bottom row, showing the smaller scale field that would
             contain high temperature plasma that constitutes the star's X-ray bright corona (top row), the larger scale field that interacts with 
             the inner disk funneling gas on to the stellar
             surface (second row), and open field lines along which a stellar wind could be launched (third row).  The star is viewed at inclination 45$^\circ$ and at phase 0.5.
             The top image has been expanded for clarity and is not on the same scale as the others.  The white points in the plot of latitude versus longitude are the footpoints of 
             the accretion funnels showing where the hot spots would be on the stellar surface, assuming an anti-symmetric field distribution with respect to the 
             stellar midplane (see section \ref{tiltlabel}).}
\label{v2247oph}
\end{figure}

Figure \ref{v2247oph} shows potential field extrapolations from the observationally
derived magnetic map of the accreting T Tauri star V2247 Oph (Donati et al 2010a).  This star has a complex magnetic field (see the map in 
Figure \ref{complexsurfacemaps})
with a weak dipole component ($<0.1\,{\rm kG}$); none-the-less the dipole component dominates on the large scale (but see section \ref{comphw}).
A source surface of $R_S\sim4.3\,R_\ast$, where $R_\ast=1.6\,{\rm R}_\odot$, comparable to the equatorial corotation radius
for V2247 Oph ($R_{co}=4.25\,R_\ast$ with $M_\ast=0.36\,{\rm M}_\odot$, $P_{rot}=3.5\,{\rm d}$) has been chosen for illustrative purposes.  The 
radius is different from that in Donati et al (2010a) as we have updated the distance estimate to V2247~Oph from $140\,{\rm pc}$ 
to $120\,{\rm pc}$ (Loinard et al 2008), which yields a different luminosity and consequently a different radius.  Assuming accretion occurs from 
a range of radii within corotation the bottom row of Figure \ref{v2247oph} shows the location of the accretion spots on the stellar surface.  The main 
spot in the visible hemisphere is arc-shaped and decreases in latitude towards the stellar equator from phase $\sim0.4$ to $\sim0.8$.  This is in excellent
agreement with the accretion spot location inferred from the excess emission map derived by Donati et al (2010a; see their figure 7). 


\subsection{Analytic models}\label{analytic}
Multipolar magnetic fields can also be handled analytically, including highly complex magnetic fields with many
high order components.  For the analysis in this work we limit ourselves to the zonal (also called axial) multipoles of arbitrary $\ell$ number
(or linear combinations of $\ell$ numbers).  Thus we are neglecting the non-axisymmetric field modes (those with $m\ne0$).  The same
assumption is made in the generalized multipolar X-wind model of Mohanty $\&$ Shu (2008) and in the initial field structures used as inputs
in current 3D MHD simulations (Romanova et al 2011; Long et al 2011; see section \ref{conclu}). Observationally the magnetic fields of accreting T
Tauri stars are dominantly axisymmetric with the exception of some low mass stars (typically below $0.5\,{\rm M}_\odot$) and older/more massive 
T Tauri stars with substantial radiative cores (those with $M_{core}/M_\ast\ge0.4$; Gregory et al 2012).  

Gregory et al (2010) derived an expression for the spherical magnetic field components $(B_r,B_\theta,B_\phi)$ exterior to a
star for a multipole of order $\ell$ (where $\ell=1,2,3,4,\ldots$ are the dipole, the quadrupole, the octupole, the hexadecapole etc),
{\setlength{\mathindent}{0pt}
\begin{eqnarray}
\mathbf{B}&=&\frac{B_\ast^{\ell,pole}}{(\ell+1)}\left(\frac{R_\ast}{r}\right)^{\ell+2}\times \nonumber \\
 &&\sum_{k=0}^N \Big\{ \frac{(-1)^k(2\ell-2k)!}{2^\ell k!(\ell-k)!(\ell-2k)!} (\hat{\boldsymbol\mu}\cdot\hat{\mathbf{r}})^{\ell-2k}\times \nonumber \\
 && \left[(2\ell-2k+1)\hat{\mathbf{r}}-(\ell-2k)(\hat{\boldsymbol\mu}\cdot\hat{\mathbf{r}})^{-1}\hat{\boldsymbol\mu} \right] \Big\},
\label{genB2}
\end{eqnarray}}
where $N=\ell/2$ or $(\ell-1)/2$ whichever is an integer, $B_\ast^{\ell,pole}$ is the polar field strength of the $\ell$-number multipole, $\mathbf{\hat{r}}$
is a unit vector along the radial direction and $\hat{\boldsymbol\mu}$ is a unit vector in the direction of the multipole moment $\boldsymbol\mu$.
Magnetic fields consisting of multipoles with many high order $\ell$-number components can be constructed by summing equation (\ref{genB2})
over the $\ell$ numbers of interest.

In the following subsections we derive expressions for the spherical field components that are written in terms of the tilt and phase of tilt of each
multipole moment explicitly.  Using these expressions we construct simulated magnetic maps for the four stars shown in Figure \ref{surfacemaps},
all of which have large scale fields that are well described by tilted dipole plus tilted octupole components.  We compare the simulated
maps to the maps derived from Zeeman-Doppler imaging which contain information about many higher order $\ell$ number 
and $m\ne0$ field modes.  We further derive generalized expressions for the stellar disk averaged longitudinal field component and demonstrate that 
reliance on this sole diagnostic provides poor constrains on the overall surface magnetic field topology (e.g. Donati $\&$ Landstreet 2009).   


\subsubsection{Tilted multipolar magnetic fields}\label{tiltlabel}
We consider an axial multipole ($m=0$) of order $\ell$ tilted by an angle $\beta$ relative to the stellar rotation axis 
and tilted towards a longitude $\psi$ where $\psi$ is related to the rotation phase $\Phi$ via
\begin{equation}
\psi=(1-\Phi) \times 360^\circ.
\end{equation}
Note that rotation phase $\Phi$ and longitude $\psi$ run backwards such that longitudes $\psi=0^\circ, 90^\circ, 180^\circ, 270^\circ$ and $360^\circ$
correspond to rotation phases $\Phi=1, 0.75, 0.5, 0.25$ and zero, and that our derivation assumes the $xz$-plane corresponds to a rotation 
phase of zero (see Appendix \ref{tiltmaths}).  The spherical field components can be derived from equation (\ref{genB2}), following 
the details of Appendix \ref{tiltmaths}, and by noting that as $\hat{\boldsymbol\mu}=\mu^{-1}(\mu_r,\mu_\theta,\mu_\phi)$ then 
$\hat{\boldsymbol\mu}\cdot\hat{\mathbf{r}}=\mu_r/\mu$,
{\setlength{\mathindent}{0pt}
\begin{eqnarray} 
&& B_r         = B_\ast^{\ell,pole} \left(\frac{R_\ast}{r}\right)^{\ell+2}\times \nonumber \\
&& \sum_{k=0}^N \left\{\frac{(-1)^k(2\ell-2k)!}{2^\ell k!(\ell-k)!(\ell-2k)!} \left(\frac{\mu_r}{\mu}\right)^{\ell-2k}\right\}\label{Brgeneral} \\
&& B_\theta = \frac{B_\ast^{\ell,pole}}{(\ell+1)} \left(\frac{R_\ast}{r}\right)^{\ell+2} \times \nonumber \\
&& \sum_{k=0}^N \left\{\frac{(-1)^{k+1}(2\ell-2k)!}{2^\ell k! (\ell-k)! (\ell-2k-1)!} \left(\frac{\mu_r}{\mu}\right)^{\ell-2k-1} \frac{\mu_\theta}{\mu}\right\}\label{Btgeneral} \\
&& B_\phi    = \frac{B_\ast^{\ell,pole}}{(\ell+1)} \left(\frac{R_\ast}{r}\right)^{\ell+2} \times \nonumber \\ 
&& \sum_{k=0}^N \left\{\frac{(-1)^{k+1}(2\ell-2k)!}{2^\ell k! (\ell-k)! (\ell-2k-1)!} \left(\frac{\mu_r}{\mu}\right)^{\ell-2k-1} \frac{\mu_\phi}{\mu}\right\}\label{Bpgeneral}.
\end{eqnarray}}
where we again note that $N=\ell/2$ or $(\ell-1)/2$ whichever is an integer, $\mu=|\boldsymbol\mu|$, and the components of the 
multipole moment $\boldsymbol\mu$ are
{\setlength{\mathindent}{0pt}
\begin{eqnarray}
\mu_r         &=& \mu\sin{\theta}\cos{\phi}\cos{\psi}\sin{\beta}+\mu\sin{\theta}\sin{\phi}\sin{\psi}\sin{\beta} \nonumber \\ 
                   &+&\mu\cos{\theta}\cos{\beta} \label{mu_r} \\
\mu_\theta &=& \mu\cos{\theta}\cos{\phi}\cos{\psi}\sin{\beta}+\mu\cos{\theta}\sin{\phi}\sin{\psi}\sin{\beta} \nonumber \\
                   &-& \mu\sin{\theta}\cos{\beta} \label{mu_theta} \\
\mu_\phi    &=& -\mu\sin{\phi}\cos{\psi}\sin{\beta}+\mu\cos{\phi}\sin{\psi}\sin{\beta} \label{mu_phi}
\end{eqnarray}}
where $\theta$ and $\phi$ are standard spherical coordinates in the reference frame of the star.
The components of a multipolar magnetosphere can then be determined by summing the appropriate field components.  
For example, a dipole plus octupole magnetosphere where the dipole (octupole) moment of polar strength $B_{dip}$ ($B_{oct}$) 
is titled by $\beta_{dip}$ ($\beta_{oct}$), relative to the stellar rotation axis, towards a rotation phase denoted by 
$\psi_{dip}$ ($\psi_{oct}$) has spherical components
\begin{eqnarray}
B_r &=& B_r(B_{dip},\beta_{dip},\psi_{dip},\ell=1) \nonumber \\
&+& B_r(B_{oct},\beta_{oct},\psi_{oct},\ell=3) \\
B_\theta &=& B_\theta(B_{dip},\beta_{dip},\psi_{dip},\ell=1) \nonumber \\ 
&+& B_\theta(B_{oct},\beta_{oct},\psi_{oct},\ell=3) \\
B_\phi &=& B_\phi(B_{dip},\beta_{dip},\psi_{dip},\ell=1) \nonumber \\
&+& B_\phi(B_{oct},\beta_{oct},\psi_{oct},\ell=3)
\end{eqnarray}
where $B_{dip}\equiv B_\ast^{1,pole}$ and $B_{oct}\equiv B_\ast^{3,pole}$.

\begin{figure*}
\centering
\includegraphics[width=85mm]{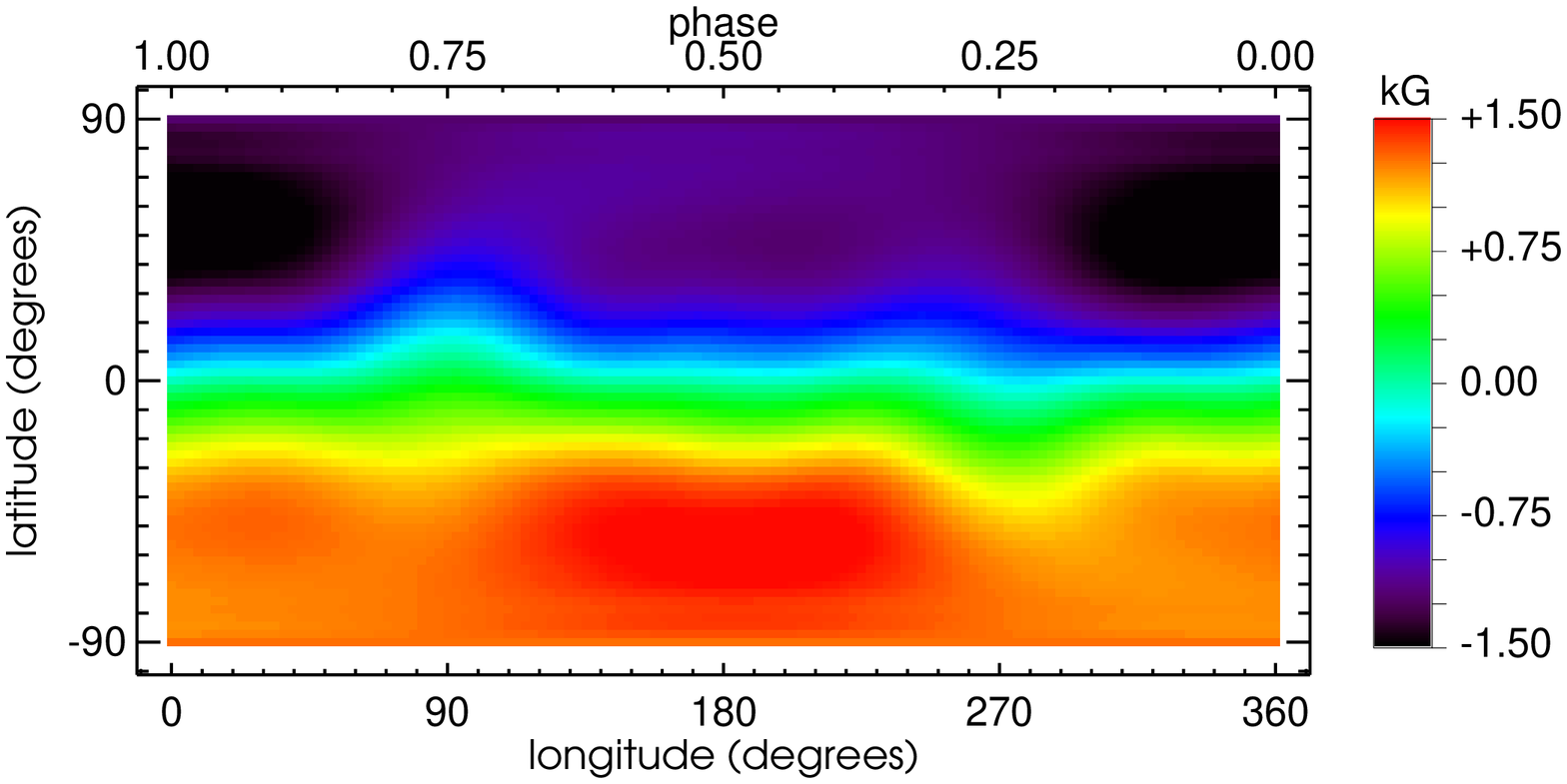}
\includegraphics[width=85mm]{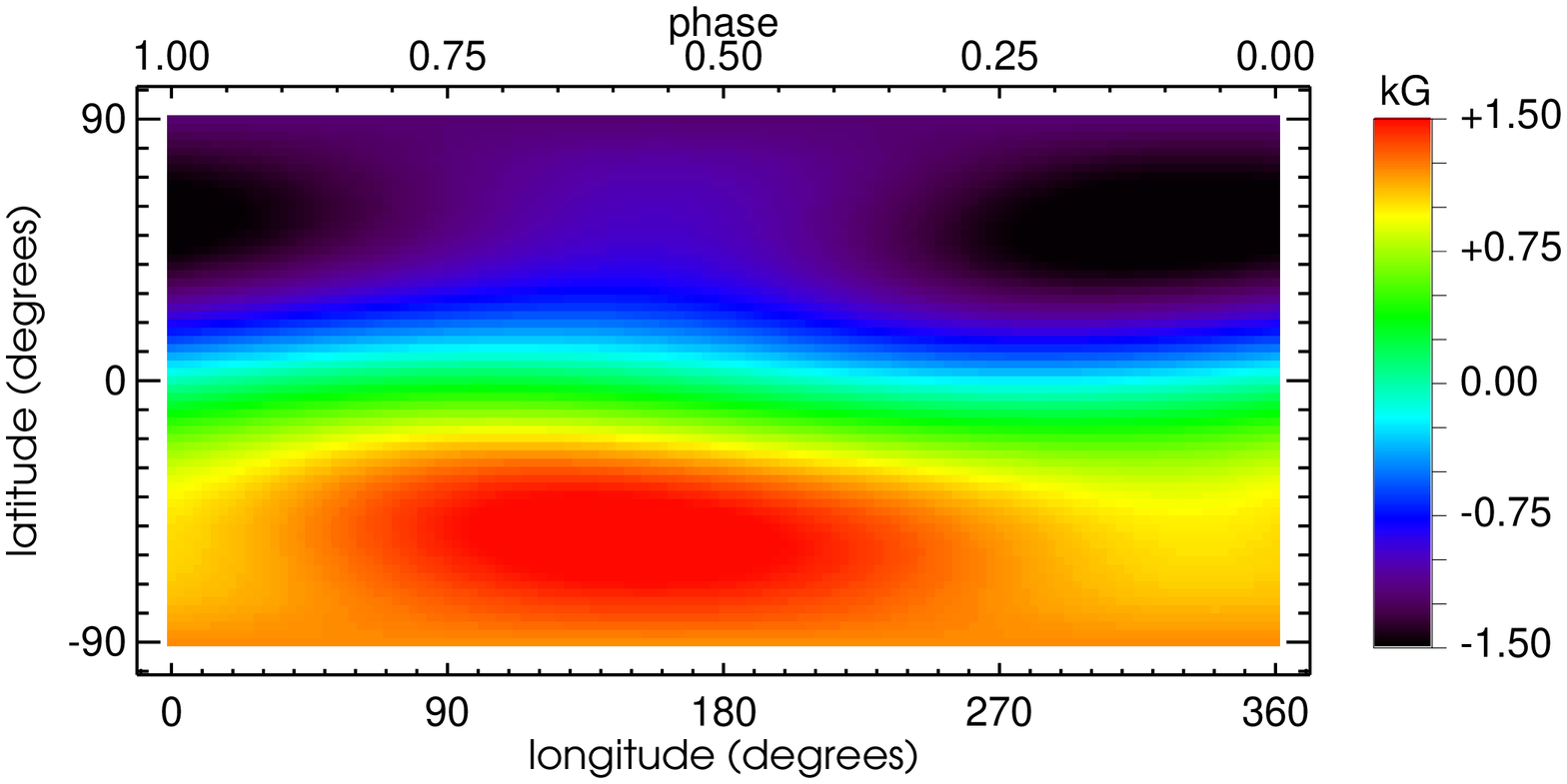} \\
\includegraphics[width=85mm]{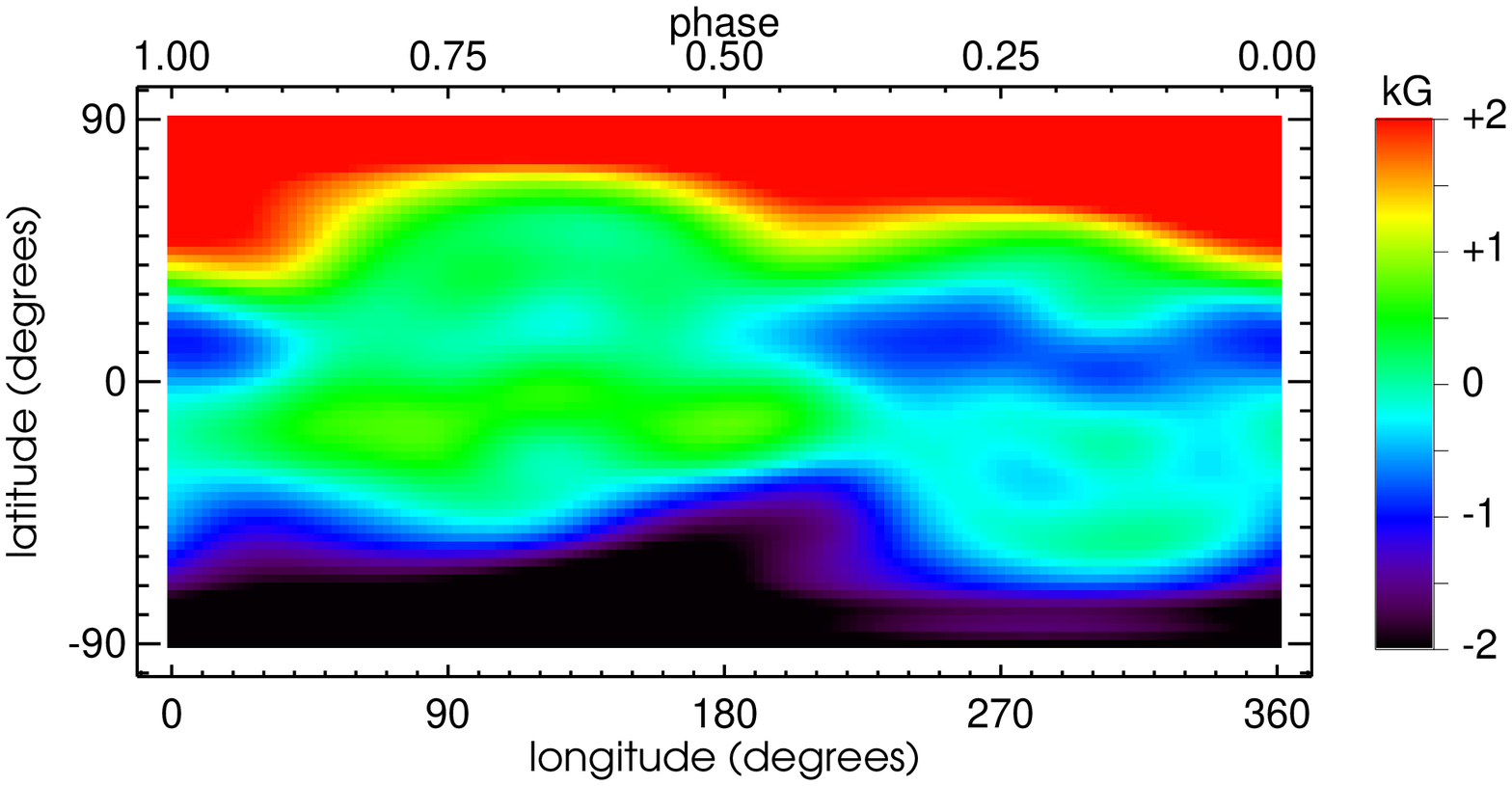}
\includegraphics[width=85mm]{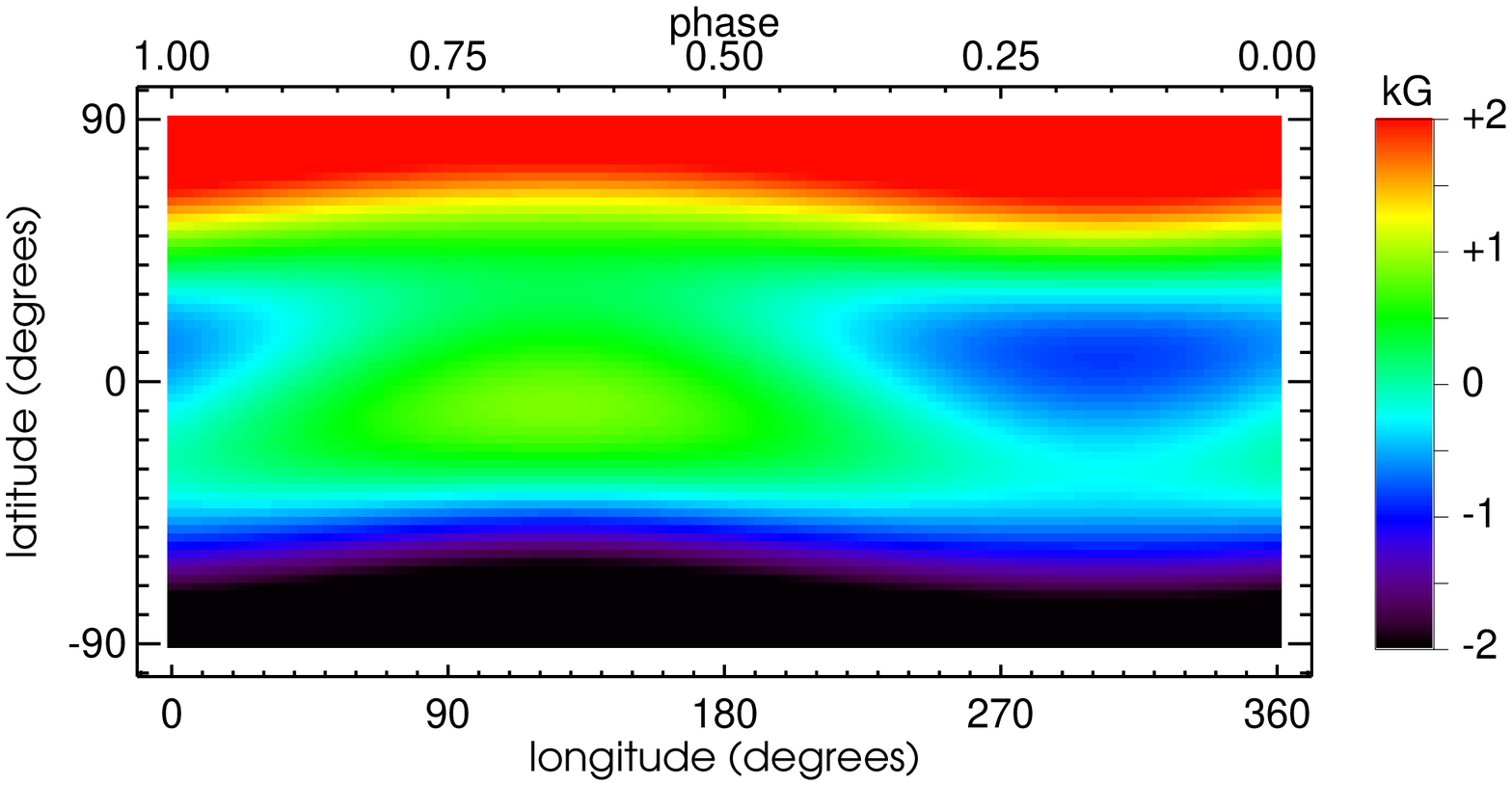} \\
\includegraphics[width=85mm]{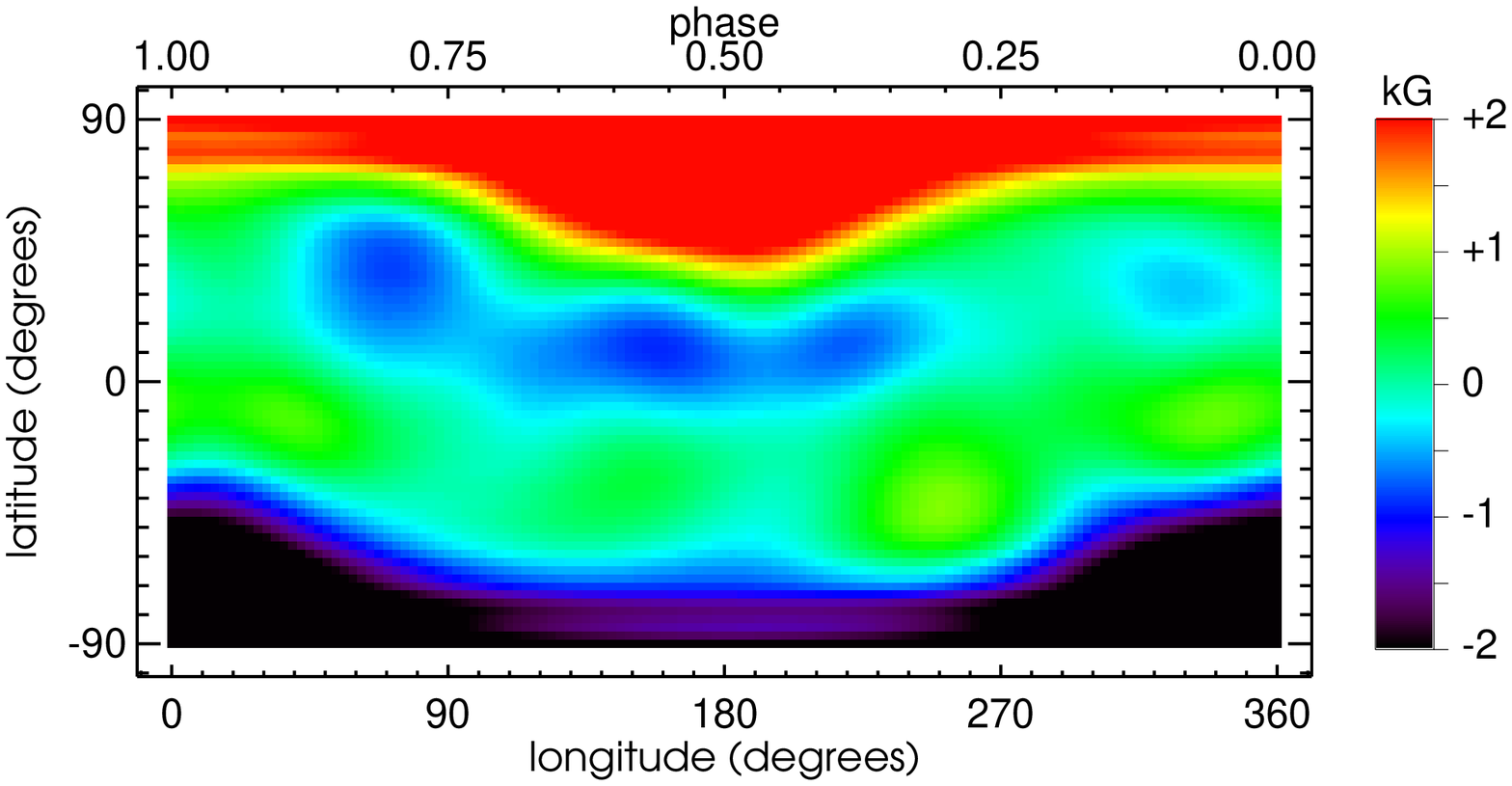}
\includegraphics[width=85mm]{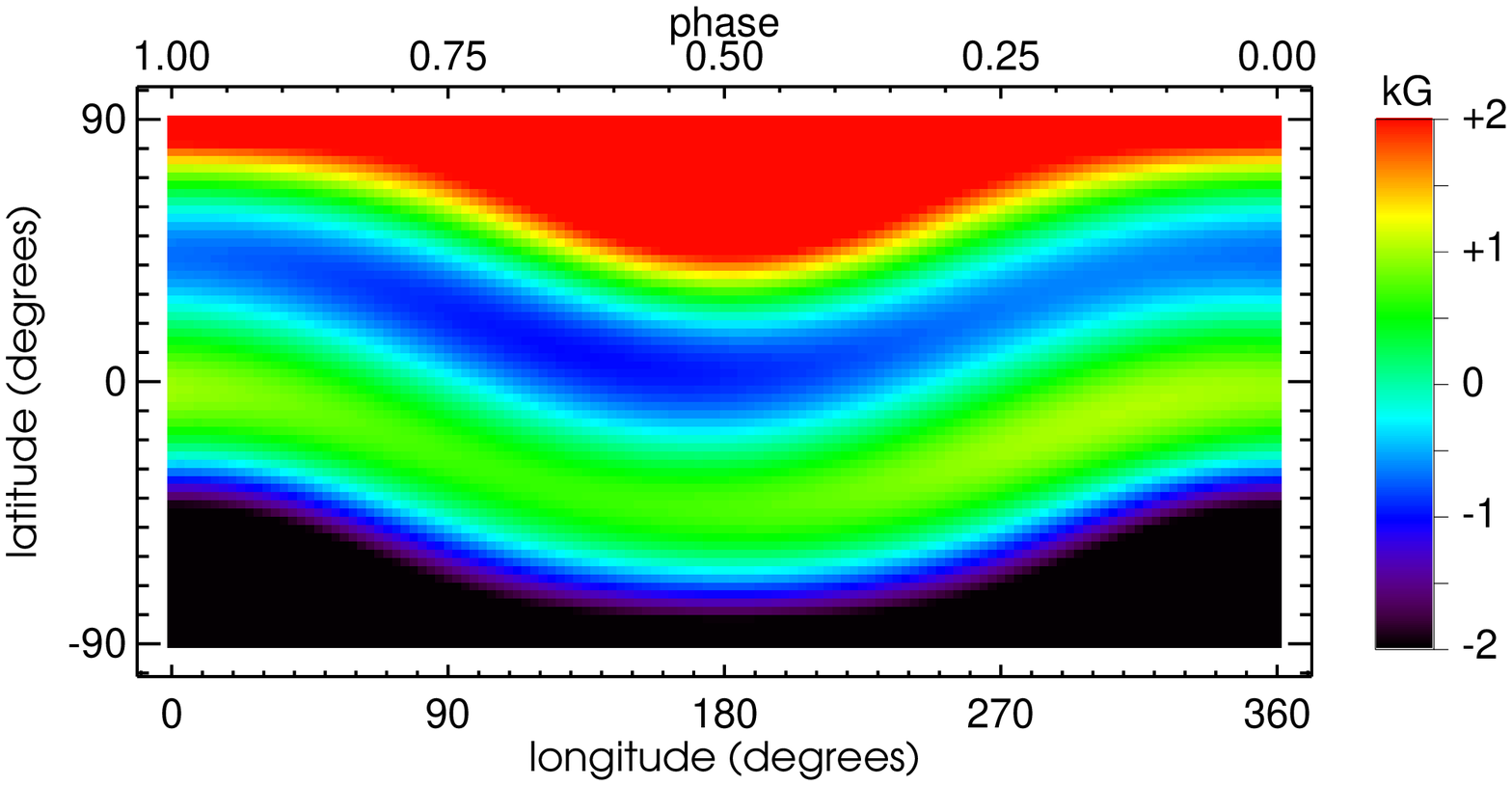} \\
\includegraphics[width=85mm]{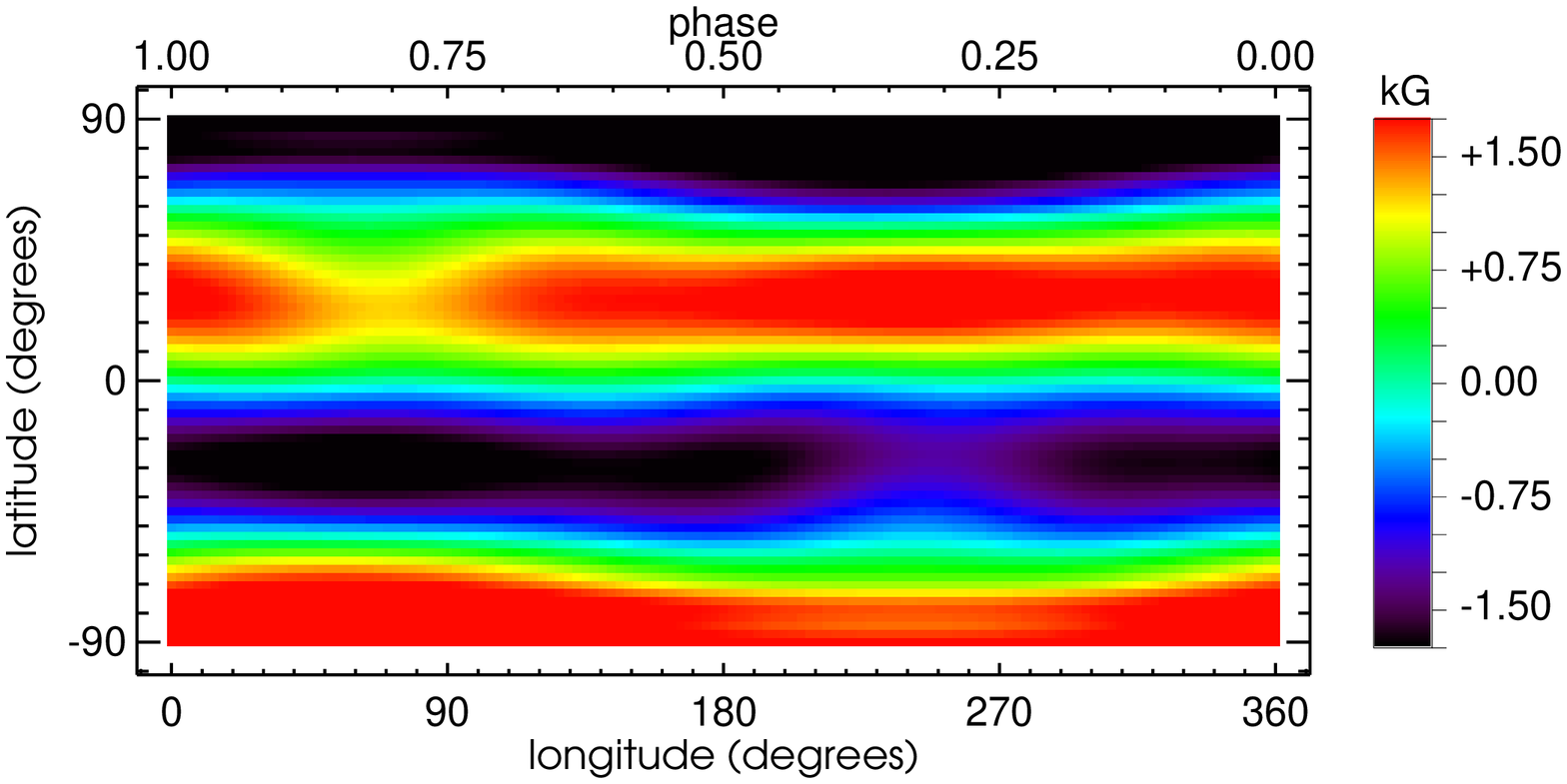}
\includegraphics[width=85mm]{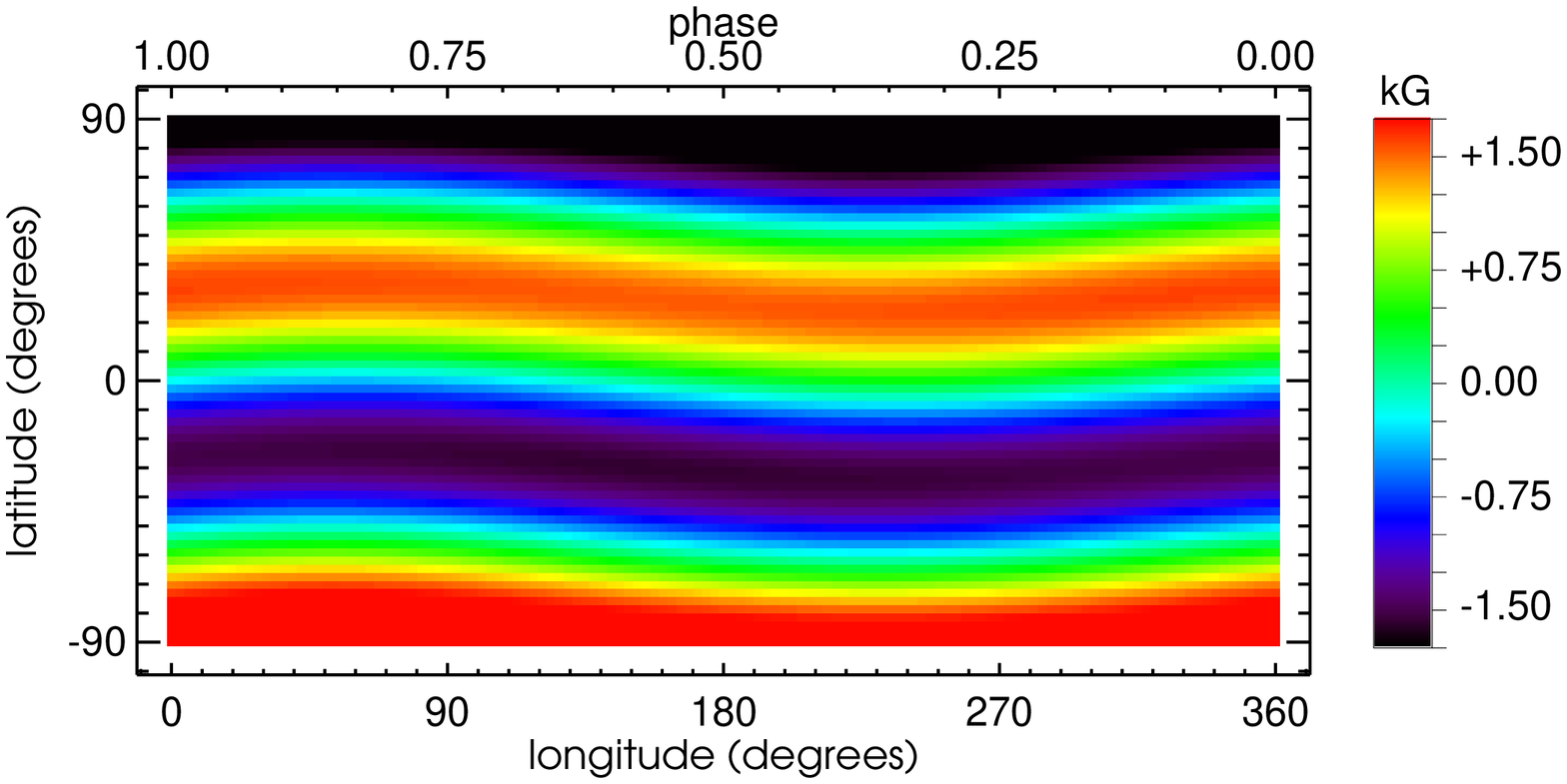} 
\caption{(left) Observationally derived magnetic maps assuming antisymmetric field configurations with respect to the stellar midplane for (top to bottom) AA Tau in January 2009, BP Tau
              in February 2006, V2129 Oph in July 2009 and TW Hya in March 2010.  (right) simulated magnetic maps constructed using tilted dipole and tilted octupole
              components with the same polar field strength, tilts, and phases of tilt as determined from the observationally derived magnetic maps as listed in Table \ref{table}.}
\label{allsimmaps}
\end{figure*}

We note that the field components derived in this section for tilted axial multipoles (those with $m=0$) also apply to sectoral multipoles 
(those with $\ell=m$) which are axial (zonal) multipoles with the moment symmetry axis tilted into the equatorial plane, i.e. perpendicular to the stellar rotation axis.

As a test of the above Figure \ref{allsimmaps} shows simulated magnetic surface maps for the four stars shown in Figure \ref{surfacemaps}.
The polar strength of the dipole and octupole components, the phase they are tilted towards, and their tilts relative to the stellar rotation axis 
match those listed in Table \ref{table}.  For comparison we also show in Figure \ref{allsimmaps} the observationally 
derived magnetic surface maps in the same Cartesian (i.e. latitude vs longitude) grid format.  The $\ell$ number of a multipole
determines the number of polarity changes in the surface field if one moves along a line of constant longitude from the north to the south
pole of the star (or vice versa).  The domination of the dipole component ($\ell=1$) in the AA Tau map and the octupole component ($\ell$=3) in
the TW Hya map is clear.

In order to produce field extrapolations some assumption has to be made about the form of the magnetic field in the $\sim$hemisphere
of the star that always remains hidden from view to an observer.  In Figure \ref{allsimmaps} we have assumed that the field is dominantly antisymmetric 
with respect to the stellar midplane, and thus the odd $\ell$ number modes dominate.  If we assume that the even $\ell$ number modes 
dominate then the resulting large scale field topologies would result in magnetospheric accretion occurring primarily on to lower, equatorial, latitudes.
As this is typically not observed, and in order to ensure that material from the disk is able to accrete into the high latitude hot spots that are observed,
we favor the antisymmetric field modes and assign the field in the hidden hemisphere accordingly.  We note that this assumption has been 
examined in detail by Johnstone et al (2010) who conclude that for accreting T Tauri stars, the field in the visible hemisphere that is reconstructed
in the magnetic maps remains largely unaffected by the assumptions of the form of the magnetic field in the hidden hemisphere.

The effect of higher order field modes is apparent in the observationally derived surface magnetic maps shown in Figure \ref{allsimmaps},
particularly for V2129 Oph and BP Tau.  The high latitude single polarity magnetic spots are stronger and of different shape in the observationally derived maps compared
to the simulated maps constructed from dipole and octupole field components only.  This is caused by the presence of higher order $\ell$ numbers in the observed
maps.  Likewise there is more azimuthal structure apparent in the observed maps; in particular within the mid-latitude negative field band on both V2129 Oph and BP Tau.
This is caused by the presence of non-axisymmetric field components (those with $m\ne0$) in the observed maps (further details can be found
in the papers where the magnetic maps have been published; Donati et al 2011a for V2129 Oph; Donati et al 2008 for BP Tau; Donati et al 2010b for AA Tau; and Donati et al 
2011b for TW Hya).  The maps derived from Zeeman-Doppler imaging thus contain more information than can be captured by the simple model consisting of tilted dipole plus 
tilted octupole field components only.  None-the-less it is apparent from Figure \ref{allsimmaps} that the simple dipole-octpole model provides excellent overall agreement
on their large scale field topologies; it is such dipole-octupole models that have been used as the input conditions in the latest generation of 3D
MHD simulations of the star-disk interaction (Romanova et al 2011; Long et al 2011).  In the following section we compare the large scale field topology derived 
via field extrapolation from the two sets of surface maps presented in Figure \ref{allsimmaps}.
  

\subsubsection{Comparison with field extrapolations}\label{comphw}
\begin{figure}
\centering
\includegraphics[width=40mm]{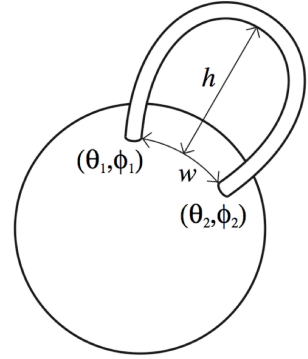}
\caption{Cartoon illustrating the definition of the height $h$ and the width $w$ of a closed field line loop with foot points at colatitudes and longitudes 
              $(\theta_1,\phi_1)$ and $(\theta_2,\phi_2)$.}
\label{hwcartoon}
\end{figure}

We have carried out field extrapolations from the magnetic surface maps derived from Zeeman-Doppler imaging and those constructed assuming 
only tilted dipole and tilted octupole field components shown in Figure \ref{allsimmaps}.  This allows for a quantitative comparison between the 
3D magnetospheric topologies derived from both sets of maps.  To compare the field topologies we calculate the height and width of the 
closed field line loops, as defined in Figure \ref{hwcartoon}.  The height $h$ is the maximum height
the loop reaches above the stellar surface, and the width $w$ is the distance between the loop foot points as measured along the segment of the great
circle that passes through the foot points on the stellar surface.  Details of how these quantities are calculated can be found in Gregory et al (2008).  

\begin{figure*}
\centering
\includegraphics[width=50mm]{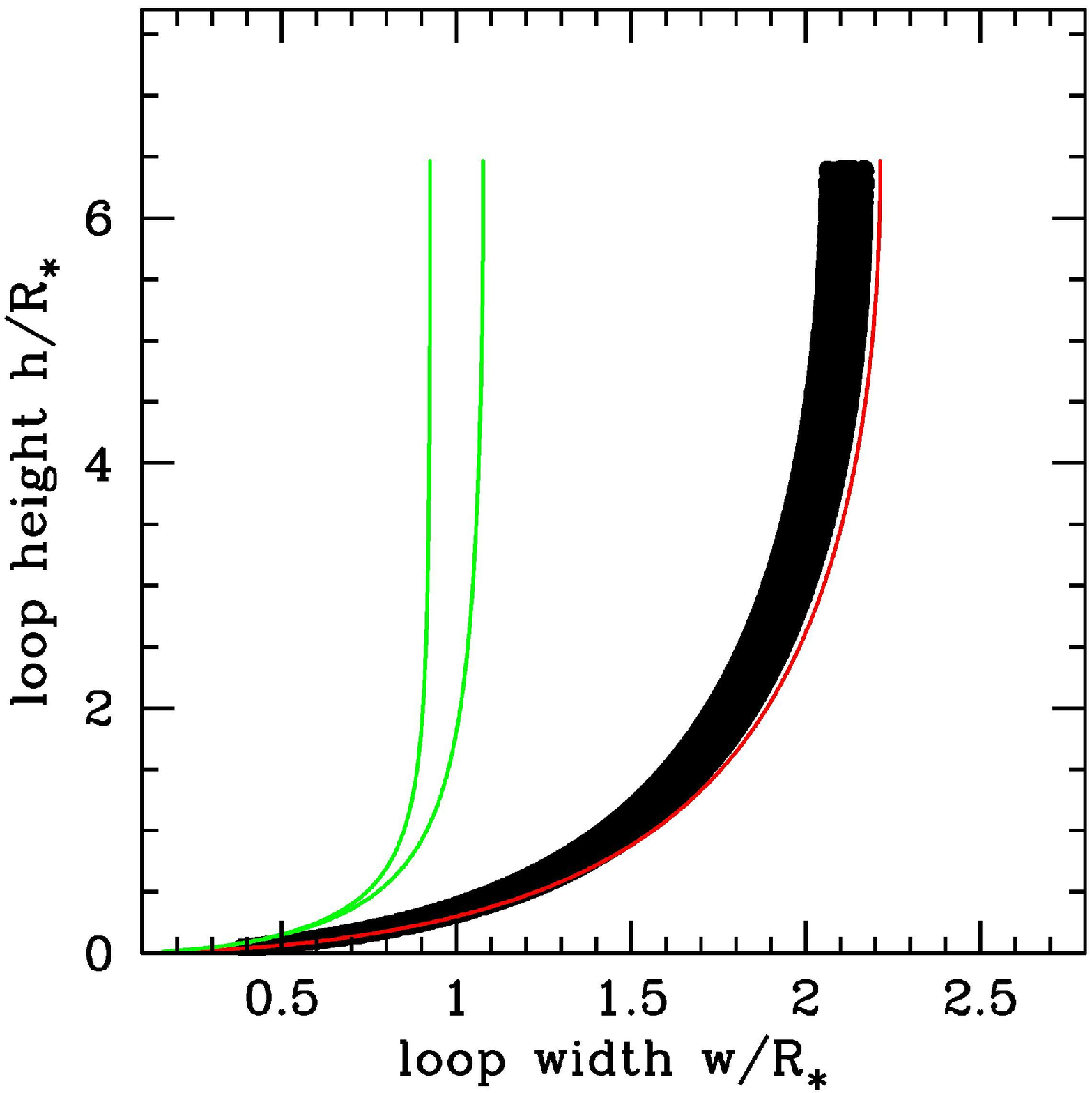}
\includegraphics[width=50mm]{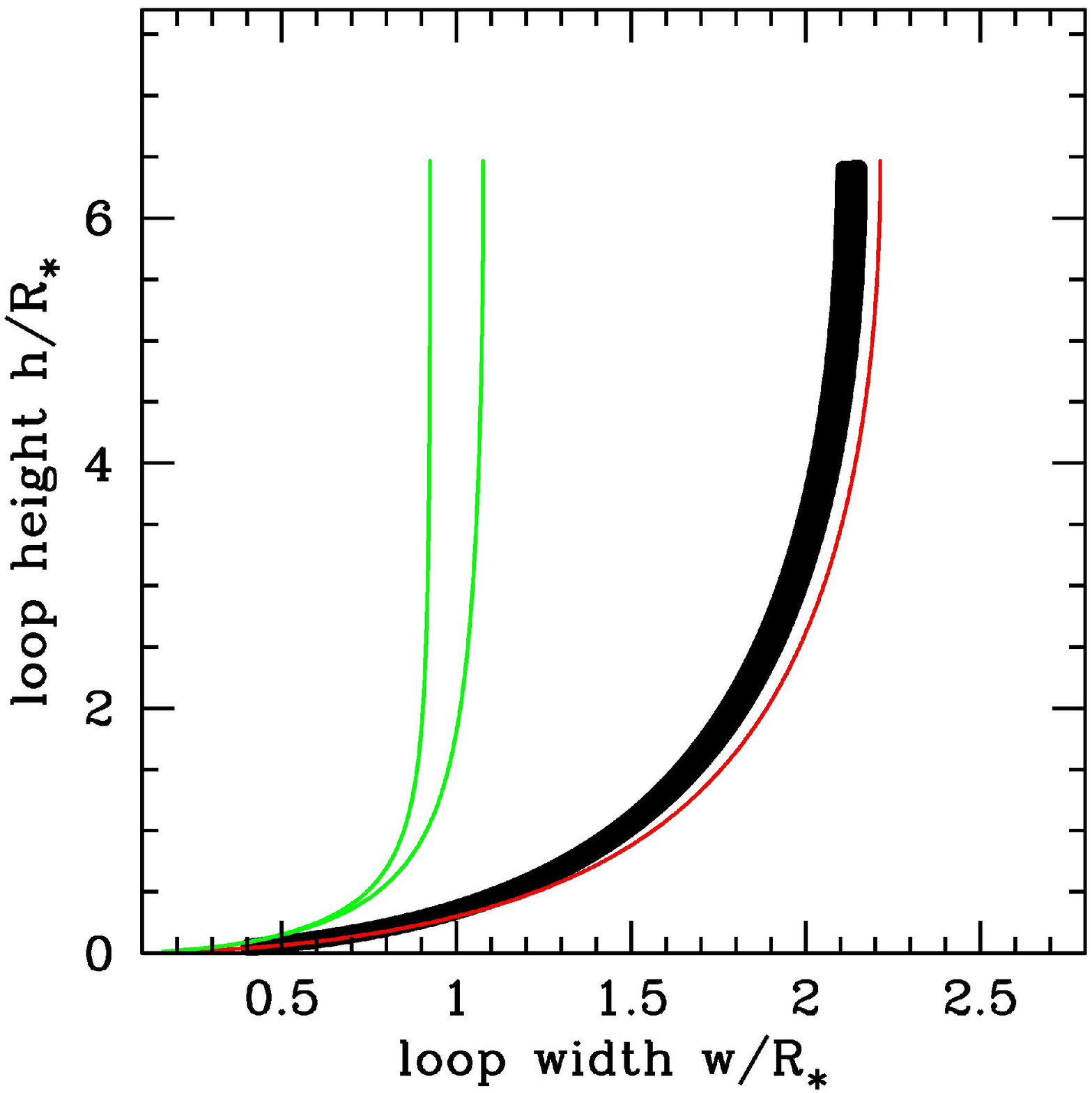} \\
\includegraphics[width=50mm]{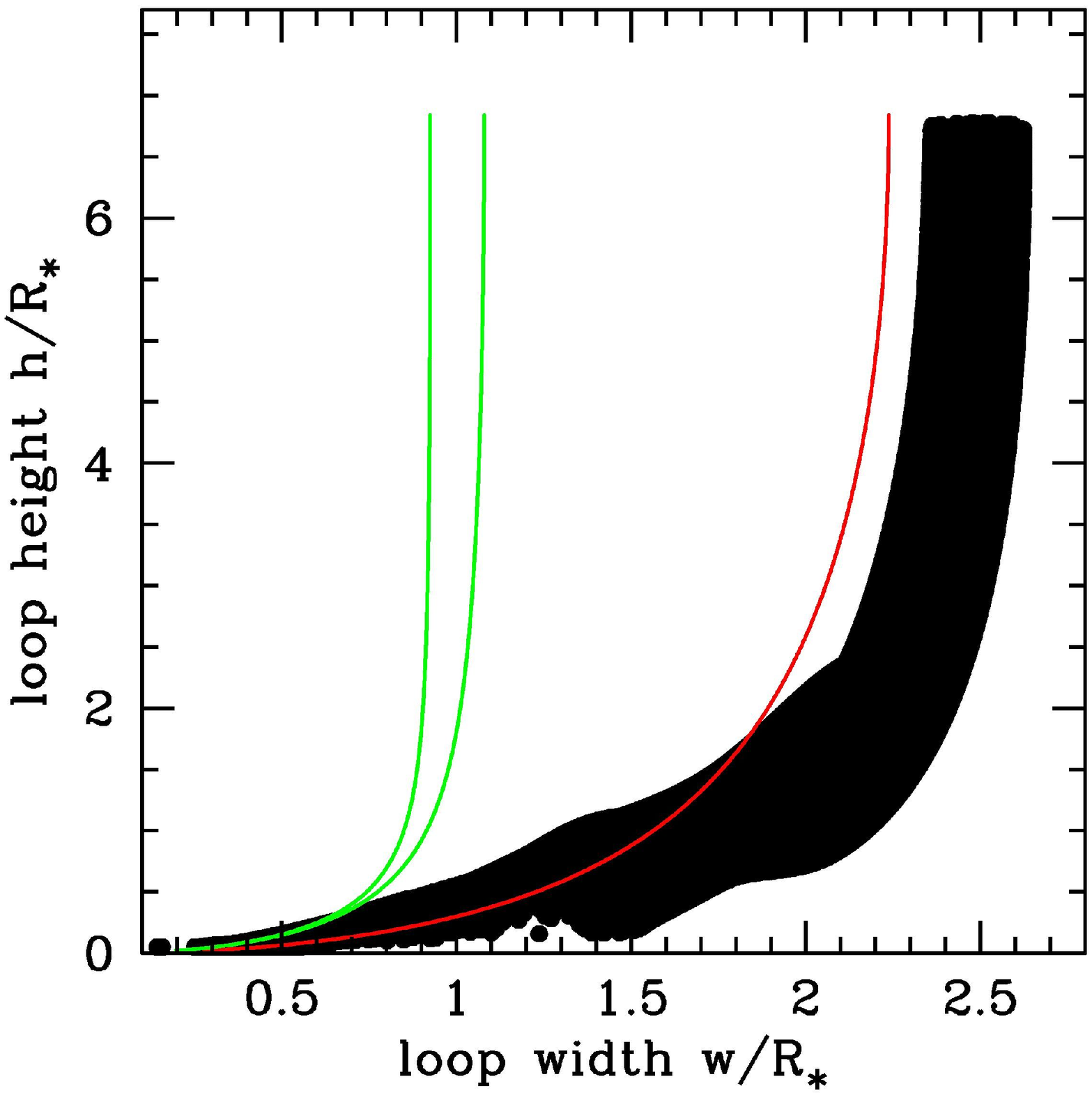}
\includegraphics[width=50mm]{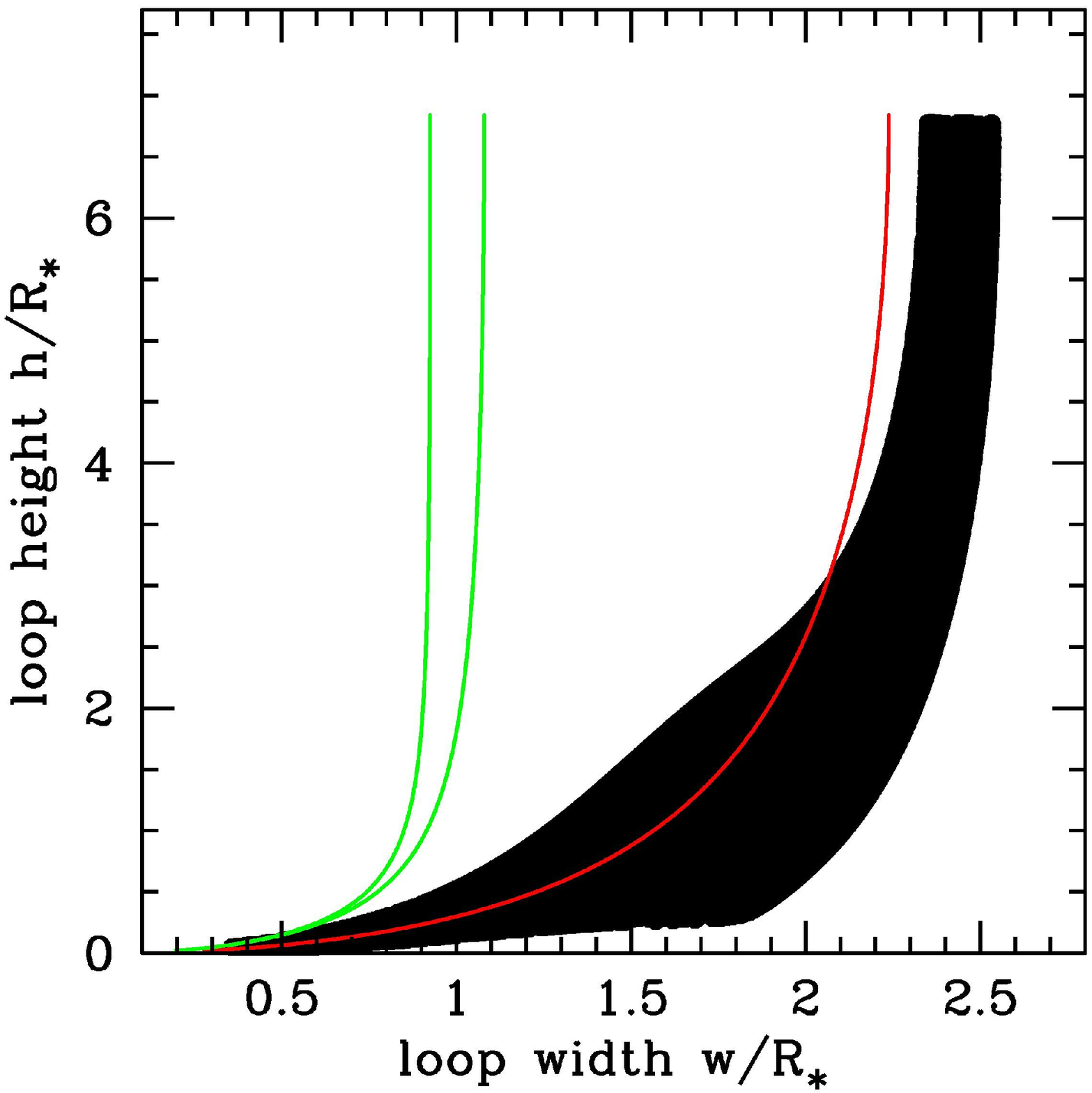} \\
\includegraphics[width=50mm]{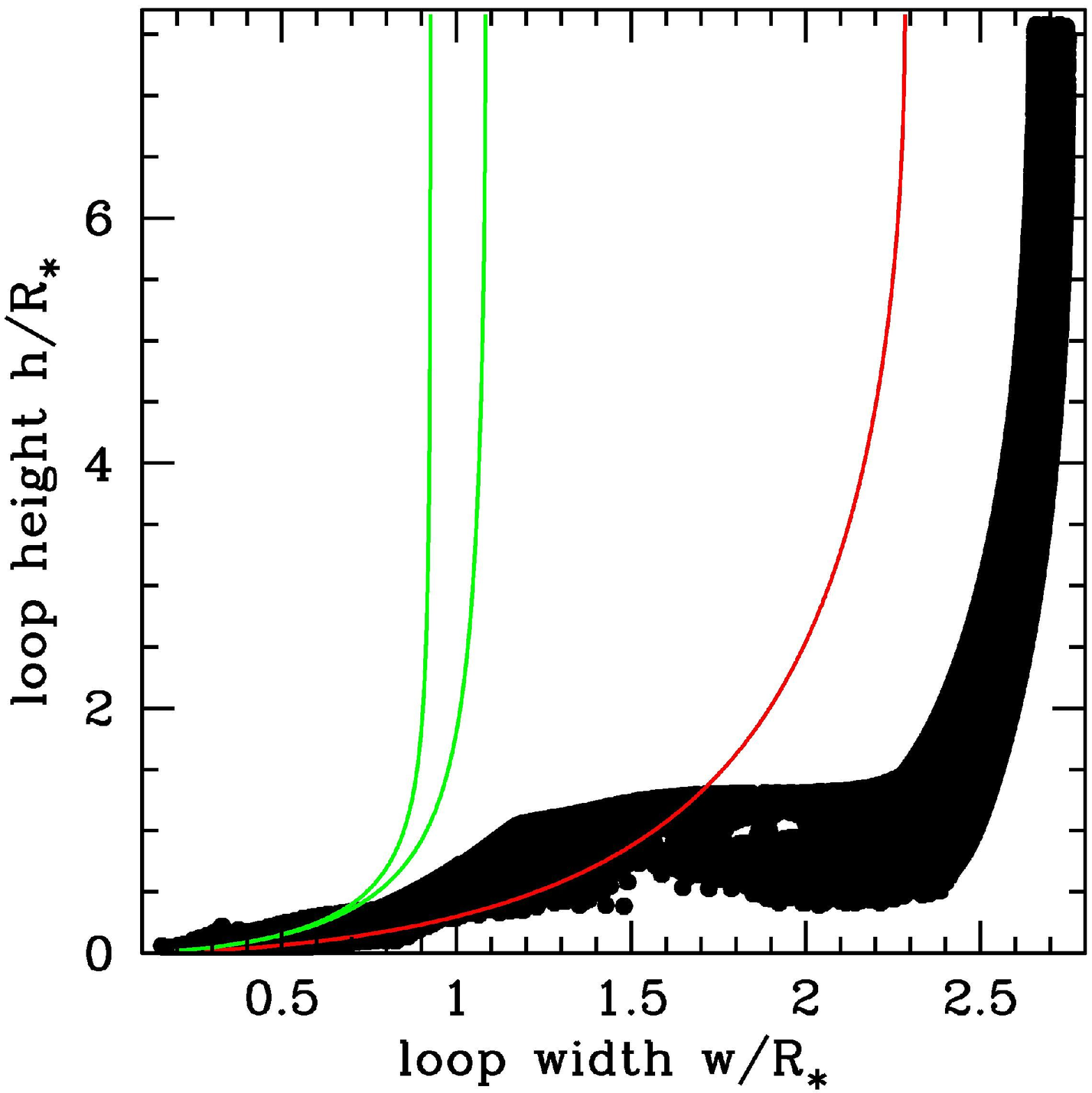}
\includegraphics[width=50mm]{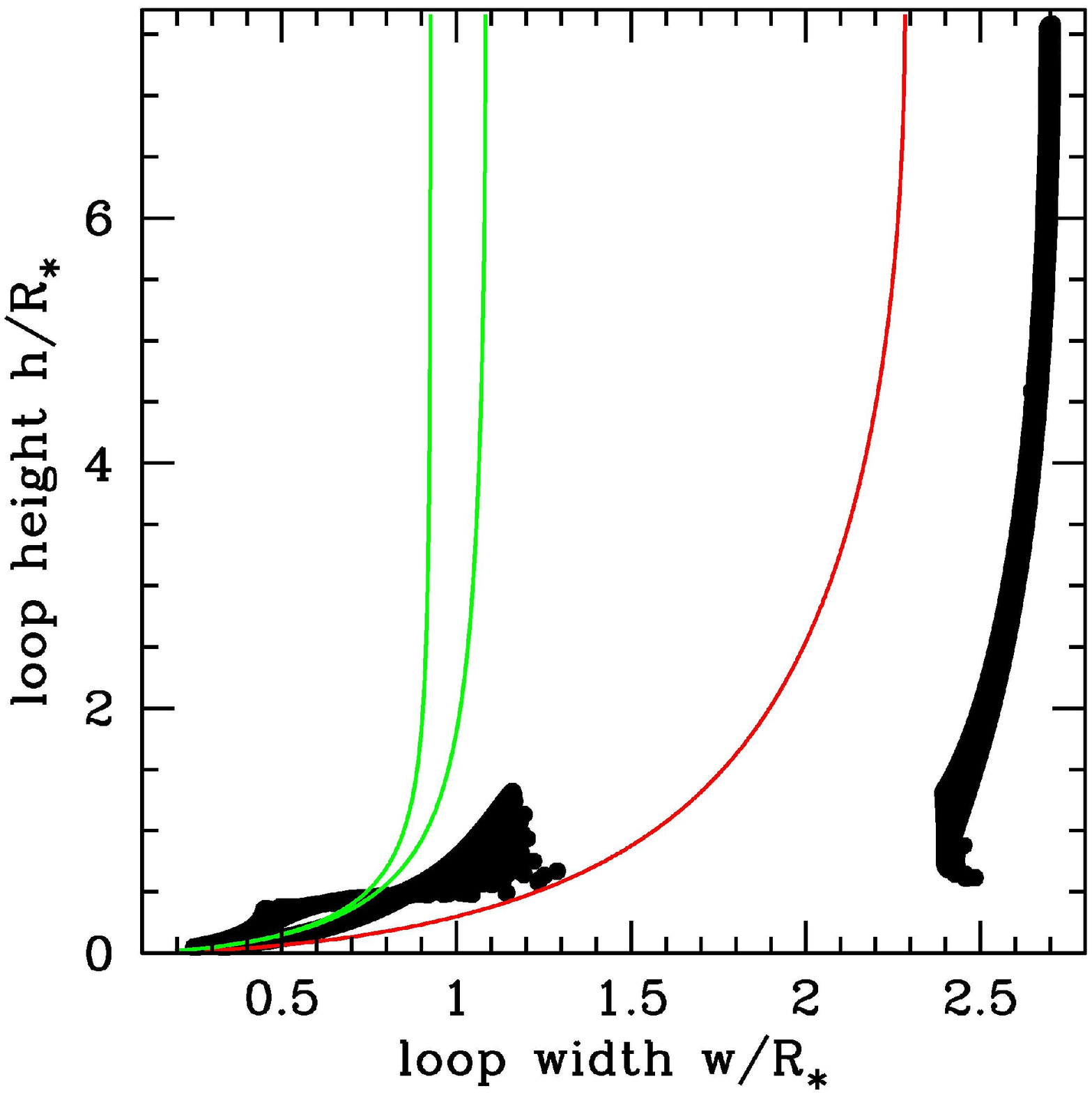} \\
\includegraphics[width=50mm]{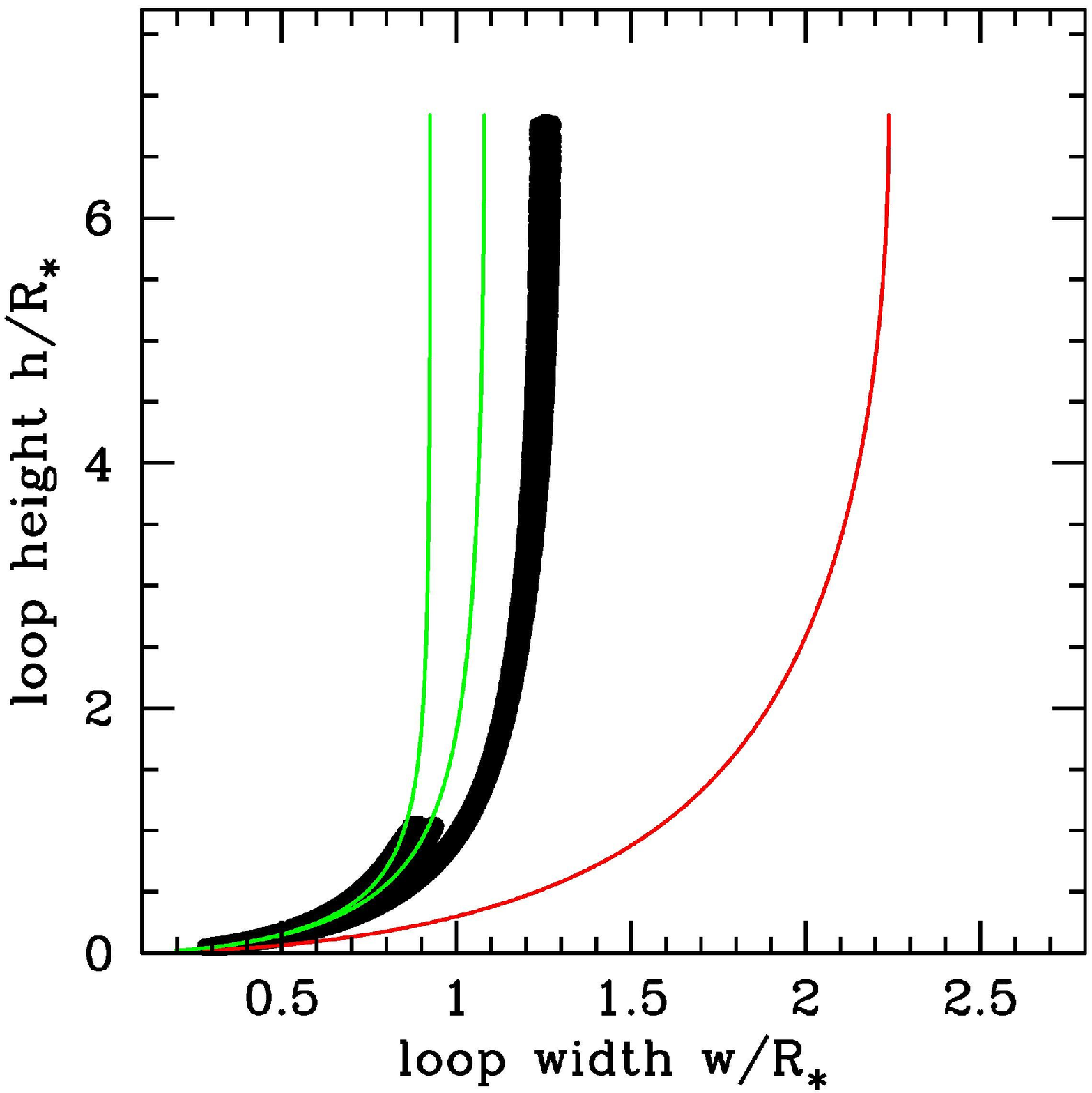}
\includegraphics[width=50mm]{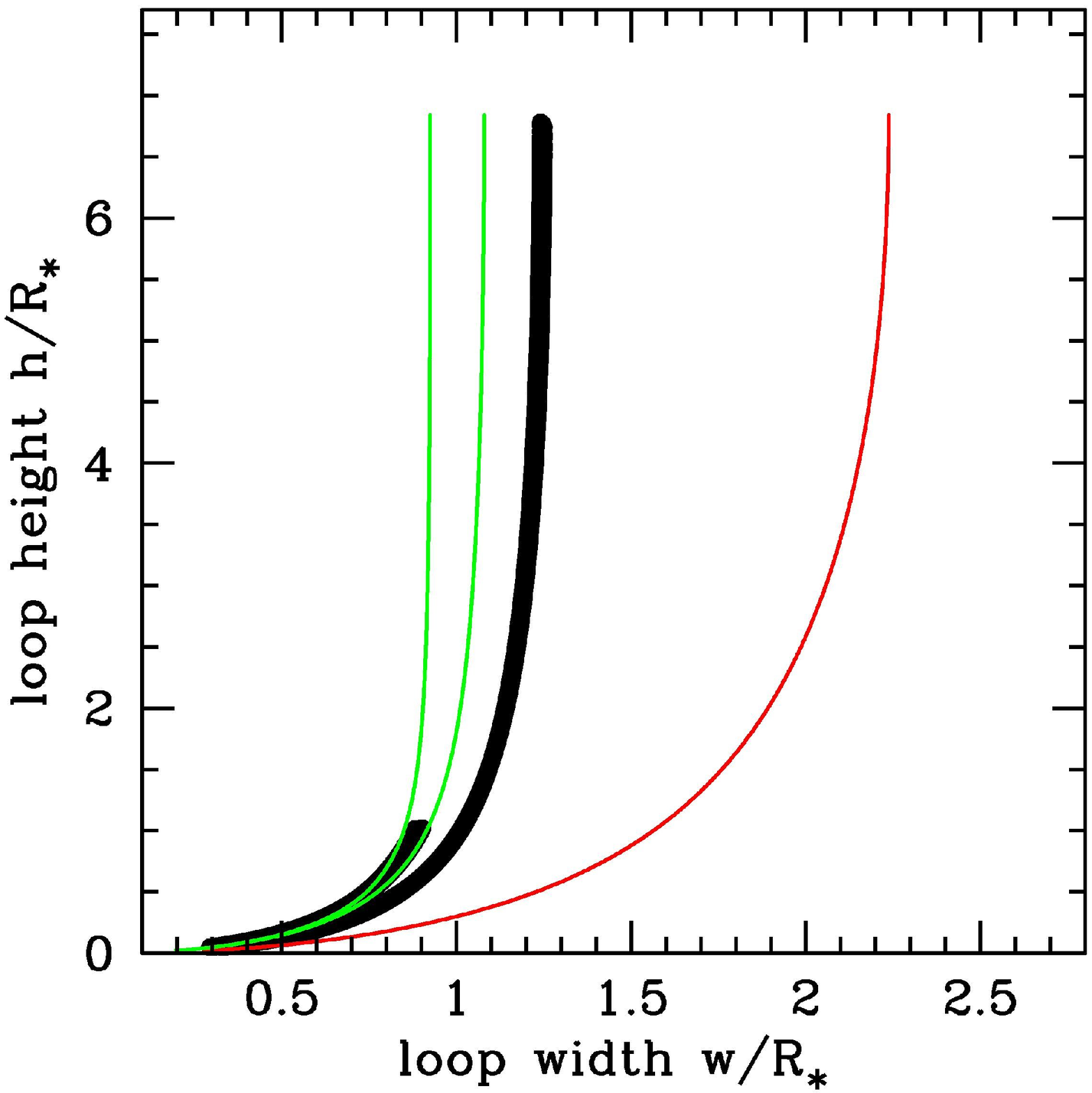}
\caption{Height $h$ versus width $w$ plots for the field lines obtained via field extrapolation from the magnetic maps in Figure \ref{allsimmaps} (black points).  Plots
             corresponding to the
             observationally derived maps are shown on the left, and those from the simulated maps assuming only tilted dipole plus tilted octupole field components are shown on the
             right for, from top to bottom, AA Tau, BP Tau, V2129 Oph and TW Hya.  On each plot the red line shows the height-width relation for an axial dipole field
             ($\ell=1$, $m=0$), and the green lines the equivalent for an axial octupole field ($\ell=3$, $m=0$).  
              }
\label{hwplots}
\end{figure*}

Figure \ref{hwplots} shows the comparison between the field extrapolations from the maps in Figure \ref{allsimmaps}.  For simplicity we have set the source
surface to be at approximately the equatorial corotation radius for each star.  We remind readers that the source surface denotes the maximum extent of the 
closed field, and thus a field line cannot have a height greater than the radius of the source surface; in this case corotation.  Larger scale field 
lines are likely torn open due to the interaction with the disk (e.g. Matt $\&$ Pudritz 2005), as has also been found in MHD simulations (e.g. Romanova et al 2011).   
The maximum width that a field line could have is less than $\pi\,R_\ast$ (loops with a larger width would be wider than the star, which is clearly unphysical).  On each plot 
in Figure \ref{hwplots} the solid red line shows the relationship between $h$ and $w$ for a pure dipole field, and the solid green lines for a pure octupole field.  There are 
two such lines for the octupole case as for an axisymmetric octupole there are three rings of closed field about the star, and the higher latitude rings are of different 
width compared to the equatorial ring (this is due to the mathematical properties of the associated Legendre functions, see Gregory 2011).  
 
It is clear from Figure \ref{hwplots} that there is good agreement between the 3D field structures obtained via field extrapolation
from the observationally derived (left column), and the simulated dipole-octupole (right column), magnetic maps.  Better
agreement is however obtained for AA Tau (top row) and TW Hya (bottom row).  These are the stars which 
have one of the dipole or octupole components clearly dominant over the other; AA Tau hosting a dominantly dipolar
magnetic field, and TW Hya a dominantly octupolar one (see Table \ref{table}).  The agreement is poorer, although not
significantly so, for BP Tau and V2129 Oph (second and third rows of Figure \ref{hwplots}), stars which have both strong
dipole and octupole field components; although for V2129 Oph/BP Tau the octupole/dipole component contains the bulk of the
magnetic energy (Donati et al 2011a, 2008).  For V2129 Oph in particular there are many field lines with large widths ($\sim2\,R_\ast$) 
but small heights ($\sim1\,R_\ast$) that are apparent in the field extrapolation from the observationally derived map (see 
Figure \ref{allsimmaps} first column third row) but which are missing from the extrapolation from the simulated map containing the 
tilted dipole and tilted octupole field components only (see Figure \ref{allsimmaps} second column third row).  These field
lines connect the strong negative field regions evident in the mid-latitude northern hemisphere to equivalent positive field regions
in the other hemisphere.  This azimuthal field structure which arises due to the presence of non-axisymmetric ($m\ne0$) field modes
can be seen in the observationally derived map.  As the simulated maps neglect the $m\ne0$ field modes for simplicity, a group of
field lines is missed in the field extrapolation from the V2129 Oph simulated map, which explains the difference seen in Figure \ref{hwplots}.
Given that these field lines are limited in height their presence does not significantly affect the structure of the larger
scale field which is interacting with the disk, and is thus unlikely to have any effect on the results of MHD modeling which use
dipole-octupole fields as inputs (see section \ref{conclu}).     

AA Tau and TW Hya have field configurations where the dipole and octupole components are close
to an anti-parallel configuration, whereas for BP Tau and V2129 Oph they are close to a parallel configuration 
(at least at the epochs listed in Table \ref{table}).  The differing field configurations are clearly apparent
in the global field structure and in the plots of field line height versus width, see Figure \ref{hwplots}.
The largest field lines, those with the greatest height, for AA Tau and TW Hya have smaller width at
the same height compared to a pure dipole, i.e. the black points in Figure \ref{hwplots} for these stars 
lie to the left of the red line that represents the dipole.  The opposite is the case for BP Tau and V2129 Oph
where the field lines with the greatest height are wider than those of a dipole (as already discussed in Gregory
et al 2008) i.e. the black points in Figure \ref{hwplots} lie to the right of the red line.

In all cases the magnitude of this effect, and the departure from a dipole field, is greater for larger values of the 
ratio of the polar field strength of the octupole to the dipole component $B_{oct}/B_{dip}$.  This can be understood
as follows.   The dipole and octupole components for each of the four stars listed in Table \ref{table} are tilted by different amounts  
towards different rotation phases.  However, as the tilts are small, let's assume that V2129 Oph and BP Tau have dipole-octupole fields
where the moments are both aligned with the stellar rotation axis and parallel to one another, $\beta_{dip}=\beta_{oct}=0^\circ$ 
(main positive poles coinciding with the visible rotation pole).  Likewise, let's assume that AA Tau and TW Hya have dipole-octupole fields where the 
moments are both aligned with the stellar rotation axis but are anti-parallel to one another, $\beta_{dip}=0^\circ$, $\beta_{oct}=180^\circ$ 
(positive/main negative pole of the dipole/octupole coinciding with the visible rotation pole; TW Hya is close to this 
configuration) or $\beta_{dip}=180^\circ$, $\beta_{oct}=0^\circ$ (negative/main positive pole of the dipole/octupole coinciding with the 
visible rotation pole; AA Tau is close to this configuration).  The field components for the simplified cases of aligned and anti-aligned dipole-octupole
fields can be derived from equations (\ref{Brgeneral}-\ref{Bpgeneral}) using (\ref{mu_r}-\ref{mu_phi}),
{\setlength{\mathindent}{-5pt}
\begin{eqnarray}  
&& B_r = B_{dip}\left(\frac{R_\ast}{r}\right)^3\cos{\theta}\cos{\beta_{dip}}+ \nonumber \\
&& \frac{1}{2}B_{oct}\left(\frac{R_\ast}{r}\right)^5(5\cos^2{\theta}\cos^2{\beta_{oct}}-3)\cos{\theta}\cos{\beta_{oct}}\label{Brdipoct} \\
&& B_\theta = \frac{1}{2}B_{dip}\left(\frac{R_\ast}{r}\right)^3\sin{\theta}\cos{\beta_{dip}}+ \nonumber \\
&& \frac{3}{8}B_{oct}\left(\frac{R_\ast}{r}\right)^5(5\cos^2{\theta}\cos^2{\beta_{oct}}-1)\sin{\theta}\cos{\beta_{oct}}\label{Btdipoct}
\end{eqnarray}}
where $B_\phi=0$ as the fields under consideration are axisymmetric, and the first/second terms in each component
are the contributions from the dipole/octupole part of the field.  Using these field components we can solve 
the differential equation describing the path of the field lines,
\begin{equation}
\frac{B_r}{{\rm d}r}=\frac{B_\theta}{r{\rm d}\theta},
\label{Bpath}
\end{equation} 
and determine the structure of the magnetic
field external to the star.  By path of the field lines we mean a function of the form $r=r(\theta)$ that describes the 
field lines in spherical coordinates.  For example, for an aligned dipole (set $B_{oct}=0$ and $\beta_{dip}=0^\circ$ in equations
(\ref{Brdipoct}) and (\ref{Btdipoct})) the result is $r={\rm const.}\sin^2\theta$ where the constant is the maximum radial extent of the 
dipole loops in the stellar midplane (a result commonly found in the literature e.g. Gregory 2011).

\begin{figure}
\centering
\includegraphics[width=75mm]{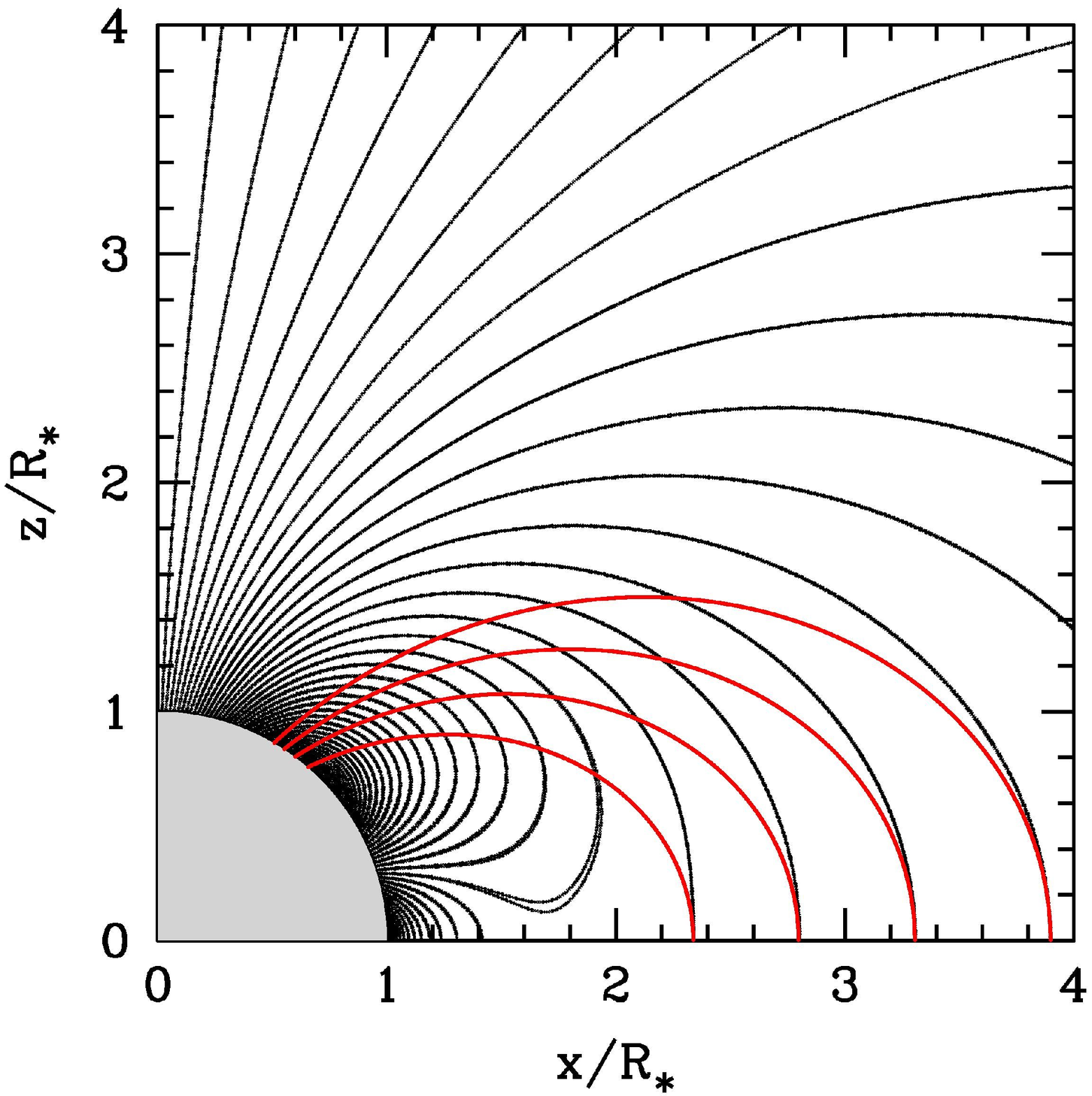} \\
\includegraphics[width=75mm]{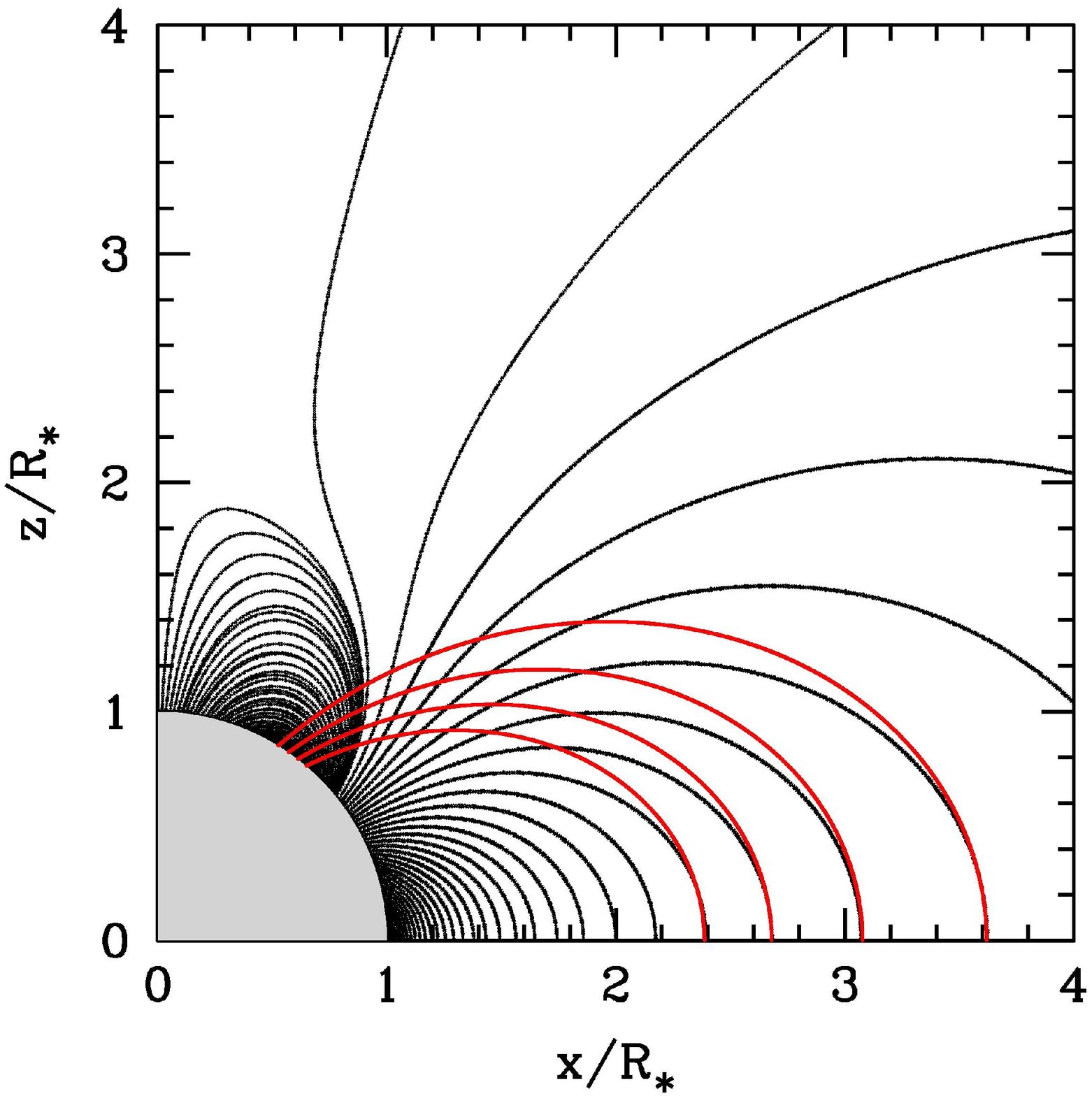}
\caption{Field lines (black) for an axisymmetric dipole-octupole field where the dipole and octupole components are parallel with $\beta_{dip}=\beta_{oct}=0^\circ$ (top panel),
             and where the dipole and octupole components are anti-parallel with $\beta_{dip}=0^\circ$ and $\beta_{oct}=180^\circ$ (bottom panel).  $B_{oct}/B_{dip}=4.0$ in both cases.
              For comparison some dipole field lines are shown in red.  The larger scale dipole-octupole field lines are squeezed by the more complex surface field.  As
              the field lines are rotationally symmetric about the $z$-axis and reflectionally symmetric in the $x$-axis only one quadrant is shown.}
\label{dipoct}
\end{figure}

For the dipole-octupole fields considered here, and in fact for any arbitrary combination of axial $\ell$ number multipoles, equation (\ref{Bpath})
has an analytic solution; however, the mathematics is just too miserable to detail here.  Equation (\ref{Bpath}), with equations (\ref{Brdipoct}) 
and (\ref{Btdipoct}), can also be solved numerically.  Figure \ref{dipoct} shows the axisymmetric dipole-octupole magnetic fields for two cases
(i) for dipole and octupole components that are aligned with the rotation axis and parallel to one another with 
$\beta_{dip}=\beta_{oct}=0^\circ$ (a configuration similar to that of V2129 Oph and BP Tau), with $B_{oct}/B_{dip}=4.0$ to more
clearly emphasize the difference between this field and a dipole; and (ii)
for dipole and octupole components that are aligned with the rotation axis and anti-parallel with $\beta_{dip}=0^\circ$ and $\beta_{oct}=180^\circ$
(a configuration similar to that of TW Hya) again with $B_{oct}/B_{dip}=4.0$.  On each plot the red lines show some dipole field lines passing through 
the stellar midplane at the same position as some of the dipole-octupole field lines (black lines).  
The field topology depends only on the ratio $B_{oct}/B_{dip}$ rather than on the polar strengths of the individual field components.   

\begin{figure}
\centering
\includegraphics[width=75mm]{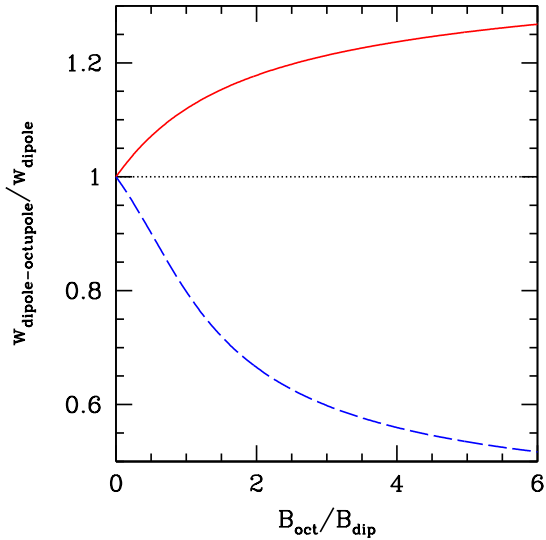} \\
\caption{The variation in the ratio of the width of a large closed field line loop for an axisymmetric dipole-octupole magnetic field $w_{dipole-octupole}$ relative to
             the width of a dipole loop $w_{dipole}$ of the same height as a function of the ratio of the polar field strength of the octupole to the dipole
             $B_{oct}/B_{dip}$.  For comparative purposes a height of $h=4\,R_\ast$ has been chosen, which corresponds to a field line threading a disk at a radius of 
             $5\,R_\ast$, a typical disk truncation radius.  The width and height are defined in Figure \ref{hwcartoon}.  The solid red line shows the case where the dipole and octupole
             components are parallel; in this senario, relevant to BP Tau and V2129 Oph, the large scale field lines are wider than the pure dipole case.  The dashed blue line 
             shows the case where the dipole and octupole and anti-parallel; in this senario, relevant to AA Tau and TW Hya, the large scale field lines are of smaller width than the 
             pure dipole case.   
	     }
\label{widthdipoct}
\end{figure}

The field line plots in Figure \ref{dipoct} provide an immediate explanation for the differences found between the extrapolated fields (the black points) and 
the dipole fields (red lines) in Figure \ref{hwplots}.  For stars like V2129 Oph and BP Tau where the dipole and octupole moments are roughly parallel, the large
scale field lines are squeezed by the octupole field line loops as we move along the large scale field lines from the stellar midplane towards the stellar
surface.  This can seen in Figure \ref{dipoct} (top panel) by comparing the dipole-octupole (black) and the dipole (red) field lines.  
In such cases the large scale field lines of the dipole-octupole fields have larger width at the stellar surface than dipole field lines of the same 
height due to the squeezing by the more complex surface field regions.  This effect becomes greater
the larger the ratio of $B_{oct}/B_{dip}$, i.e. the greater the influence of the octupole over the dipole component.  This be seen in Figure \ref{widthdipoct}
where, for a field line of a fixed height, we compare the width of the dipole-octupole field line relative to that of a dipole field line as we vary the 
ratio $B_{oct}/B_{dip}$.  As V2129 Oph has a larger $B_{oct}/B_{dip}$
ratio than BP Tau this squeezing effect is greater for V2129 Oph, and thus the large scale field lines of a given height are even wider than those of a dipole
for V2129 Oph than they are for BP Tau (see Figure \ref{hwplots} second and third rows and Figure \ref{widthdipoct}).  For field topologies like AA Tau and TW Hya, where the dipole and 
octupole moments are close to anti-parallel, the large scale field lines are distorted in a different way close to the stellar surface, see Figure \ref{dipoct}
(bottom panel).  In such cases the large scale field lines have smaller width at the stellar surface than dipole field lines of the same height, explaining the difference
found in Figure \ref{hwplots} for AA Tau and TW Hya (first and fourth rows) between the extrapolated and dipole fields.  As TW Hya has the largest $B_{oct}/B_{dip}$
value the effect is greatest for this star, but is (almost) negligible for the dominantly dipolar magnetic field of AA Tau (the field line plot using the AA Tau $B_{oct}/B_{dip}$
ratio shows very little difference from a pure dipole).  The results of this section demonstrate how
an analytic model can be used to further our understanding of the results obtained from numerical (field extrapolation) models.

The effect of the large scale field structure being distorted close to the stellar surface by the complex surface field regions is important for models of 
accretion flow along the field lines on to the stellar surface (Adams $\&$ Gregory 2011).  Not only is the hot spot location a sensitive function of the magnetic 
field topology (e.g. Gregory et al 2005, 2006a; Mohanty $\&$ Shu 2008), but the pre-shock density of gas can be up to an order of magnitude larger than that predicted 
from dipolar accretion models (Gregory et al 2007).  This is straightforward to understand physically if we assume that accretion occurs from the stellar midplane (e.g.
imagine a flat disk in the equatorial plane of Figure \ref{dipoct}).  Assuming mass conservation then the mass flux through the cross sectional area $A$ of the 
accretion column is constant, i.e. $\rho v A={\rm const.}$ where $\rho$ and $v$ are the flow density and velocity, and the constant is the mass accretion rate.  
The infall speed for gas coupled to the magnetic field is roughly the same for both accretion along dipole, and non-dipolar field lines, since the infall velocity is essentially 
free-fall and determined by the depth of the gravitational potential well of the star.  However, the cross-sectional area of the dipole-octupole field lines becomes smaller
than that of the dipole case (see Figure \ref{dipoct}), and therefore to ensure that the mass flux remains constant, the flow density must increase by roughly the same factor
that the cross sectional area reduces.    


\subsubsection{Longitudinal field component}
When high-resolution Stokes V (circular polarization) data is available detailed modeling of the profiles and their 
rotational modulation provides the best information about stellar magnetic field topologies, and allows magnetic
maps to be constructed.  However, estimates of the stellar disk averaged longitudinal field component 
(also called the transverse or line-of-sight field component) $B_{lon}$ and how this varies with 
rotation phase $\Phi$ is commonly found in the literature.  The longitudinal field component is an integrated quantity
over the visible disk of the star, and thus is often a poor diagnostic of the surface field topology due to the flux cancellation effect.  
If the stellar surface was covered in equal amounts of positive and negative field the measured stellar disk averaged 
longitudinal field component would be zero, yet in such cases the Stokes V profile itself may show a clear Zeeman signature
(e.g. see the longitudinal field curve from the LSD average photospheric absorption line and the Stokes V profile for V2129 Oph around phase 0.78 in 
July 2009; Donati et al 2011a).

Large scale magnetic topologies have been probed using the oblique rotator model, where 
a dipole field is assumed to be tilted by an angle $\beta_{dip}$ relative to the stellar rotation axis
with a stellar inclination of $i$.  The longitudinal field component thus varies with rotation phase, reaching 
a maximum when the star is viewed at a rotation phase $\Phi$ that corresponds to the phase that the dipole 
component is tilted towards $\Phi_{dip}$ (assuming it is the positive pole of the dipole in the visible hemisphere).
Here we extend our analytic work by deriving an expression for the longitudinal magnetic field component for 
dipole-octupole magnetic fields.  Even for such simple field structures, and where the octupole component is stronger than 
the dipole component, we demonstrate below that the phase modulated longitudinal field curve is dominated by 
the dipole component (although the polar field strength and tilt of this component is also poorly constrained by 
the longitudinal field curve once cool spots, neglected in this simple model, are accounted for - see below), and 
therefore provides poor constraints on the stellar field topology.  Detailed modeling 
of rotationally modulated Zeeman signatures, when high resolution spectropolarimetric data is available, 
provides better and more detailed constraints on stellar magnetic topologies (see Donati $\&$ Landstreet 2009 
for further details on the limitations of the longitudinal magnetic field component).

Although the expression for the longitudinal field component for an inclined dipole magnetosphere is commonly 
found in the literature (e.g. Preston 1967), we cannot find explicit expressions for a tilted octupole magnetosphere,
although we note that Bagnulo et al (1996) did use a combination of a tilted dipole, a tilted quadrupole and a tilted
octupole in their work.  Assuming a linear limb darkening law the stellar disk averaged longitudinal magnetic field component for a 
dipole-octupole field as a function of rotation phase is (see Appendix \ref{long} for a derivation of this result),
{\setlength{\mathindent}{-10pt}
\begin{eqnarray}
&& B_{lon}(\Phi)= \frac{1}{20}B_{dip} \left(\frac{15+u}{3-u}\right) \times \nonumber \\
&& \left(\cos{\beta_{dip}}\cos{i}+\sin{\beta_{dip}}\sin{i}\cos{[2\pi(\Phi-\Phi_{dip})]}\right) \nonumber \\
&&+ \frac{1}{16}B_{oct} \left(\frac{u-1}{3-u}\right) \times \nonumber \\ 
&&\Big\{5(\cos{\beta_{oct}}\cos{i}+\sin{\beta_{oct}}\sin{i}\cos{[2\pi(\Phi-\Phi_{oct})]})^3- \nonumber \\   
&& 3(\cos{\beta_{oct}}\cos{i}+\sin{\beta_{oct}}\sin{i}\cos{[2\pi(\Phi-\Phi_{oct})]})\Big\} \label{blondipoct},
\end{eqnarray}}
where $B_{dip}$ is the polar strength of the dipole component, $\beta_{dip}$ the tilt of the positive pole of the dipole relative to the visible 
stellar rotation pole, $\Phi_{dip}$ is the rotation phase that the dipole moment is tilted towards, with $B_{oct}, \beta_{oct}$ and $\Phi_{oct}$ 
the same respective quantities for the octupole field component.  $\Phi$ is the rotation phase, and $u$ is the linear limb darkening coefficient. 
The stellar inclination is $i$, such that the observer's reference frame $(x_{obs},y_{obs},z_{obs})$ has $z$-axis parallel to the line of sight 
i.e. we can transform from the stellar reference frame $(x,y,z)$ to the observer's frame $(x_{obs},y_{obs},z_{obs})$, which share the stellar 
center as the coordinate origin, by rotating the stellar frame counter-clockwise by an angle $i$ when looking down the $y$-axis towards 
the origin.   

\begin{figure}
\centering
\includegraphics[width=75mm]{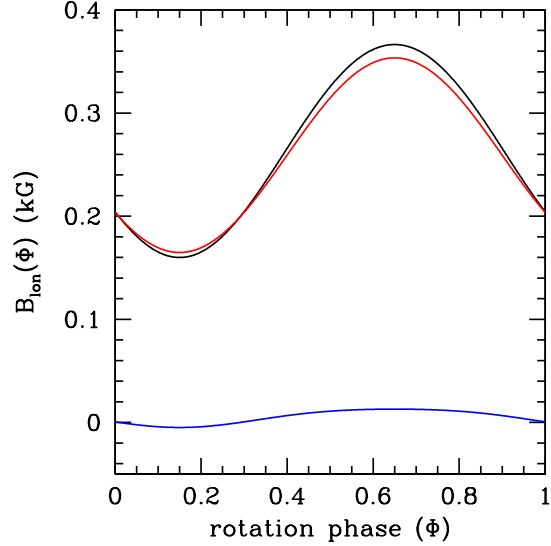} \\
\caption{The stellar disk averaged longitudinal magnetic field component $B_{lon}$ as a function of rotation phase $\Phi$ for a star with 
              a tilted dipole plus tilted octupole magnetic field with tilts, phases of tilt, and polar field strength of the components matching those
              determined from the observationally derived magnetic map of BP Tau, see Table \ref{table}.  $B_{lon}$ (black line) is calculated
              from equation (\ref{blondipoct}) assuming a linear limb darkening coefficient of $u=0.6$ and a stellar inclination of 45$^\circ$.  The
              red/blue line shows the contribution to $B_{lon}$ from the dipole/octupole component of the field.  This simple model ignores
              surface dark spots, which should be accounted for, see the text for details.}
\label{blon}
\end{figure}

In Figure \ref{blon} we illustrate the longitudinal field component as a function of rotation phase for a star with a dipole-octupole field with
the same parameters as BP Tau listed in Table \ref{table}.  It is clear that the longitudinal field curve is dominated by the dipole
component of the magnetic field even though the octupole field component is stronger.  Observationally the octupole component may easily be missed
by analyzing the longitudinal field curve alone once a limited phase coverage and errors are accounted for, despite being easily recoverable if a full analysis 
is made of the rotational modulation of the Stoke V signal, as is done in Zeeman-Doppler imaging and in the construction of stellar magnetic maps.  
The situation is worse for higher order field components.  Furthermore, even though the longitudinal field curve is only weakly sensitive to the high
order field components in the simple model presented here, it cannot be used as a probe of the large-scale dipole component.  In this simple analytic
dipole-octupole field model we have not accounted for the presence of surface cool (dark) spots.  Brightness maps are produced
as part of the magnetic imaging process, and for many accreting T Tauri stars the main magnetic pole often coincides with
large dark spots (e.g. for BP Tau; Donati et al 2008b).  Dark spots coinciding with the main magnetic pole will significantly change the respective contributions 
of the dipole and octupole field components to the longitudinal field curve, with the octupole producing a comparably larger contribution.  
If the cool spot is off-set from the main magnetic pole then the tilt of the dipole component $\beta_{dip}$ that can be derived by    
fitting equation (\ref{blondipoct}) to an observed longitudinal field curve will likely be significantly off.

If the longitudinal field curve constructed from a LSD average photospheric line was probing the 
large scale dipole component of a stellar magnetic field, then the variations of $B_{lon}$ with phase would be mostly sinusoidal; observationally
this is not usually the case.  For V2129 Oph the departure from a simple sinusoidal longitudinal field curve is obvious, see the plot in figure 2 (bottom left panel)
in Donati et al (2011a).  For AA Tau the longitudinal field component measured from the LSD average photospheric absorption line is almost featureless, and yet 
detailed modeling of the full Stokes V profile reveals a strong dipole component of polar strength $\sim2\,{\rm kG}$.  Longitudinal field 
components therefore provide poor observational constrains on T Tauri magnetic field topologies even for the dipole component itself.

Although strong and rotationally modulated longitudinal field components are often measured in accretion related emission
lines (e.g. Valenti $\&$ Johns-Krull 2004; Donati et al 2008b), there is another reason why fitting an oblique dipole rotator model to the longitudinal field curve 
may yield polar field strengths for the dipole component that are significantly discrepant from the true value.  If we use V2129 Oph in July 2009 as
a specific example, the dipole component recovered from a spherical harmonic decomposition of the magnetic map suggests a dipole
component of $\sim1\,{\rm kG}$, but the bulk of the material arrives at the star into a single positive field spot of $\sim4\,{\rm kG}$
(Donati et al 2011a). Thus the analysis of the longitudinal field component measured in accretion related emission lines which are forming close to 
the accretion shock would yield discrepant values for the dipole component (on V2129 Oph higher order field components contribute to the 
strong field region at the base of the accretion funnel). Indeed Valenti $\&$ Johns-Krull (2004) find that their longitudinal field components measured 
in the HeI 5876{\AA } emission line are rotationally modulated, but well fitted by a simple model where the accreting material lands in a single polarity 
spot on the stellar surface.  This is found in Zeeman-Doppler imaging studies for many accreting T Tauri stars (e.g. Donati et al 2008b; Donati
et al 2011a).  Analysis of the longitudinal field curve alone (whether derived from accretion related emission lines or from a LSD 
average photospheric absorption line) therefore provides poor constrains on the field topology and the polar strength of the dipole
component for accreting T Tauri stars.  A full analysis of variations in the Stokes V profile for both accretion related emission lines, and photospheric
absorption lines, is clearly desirable whenever high resolution spectra are available to provide a consistent description of the large-scale magnetic field
topology.


\section{Discussion and conclusions}\label{conclu}
In this paper we have discussed the numerical field extrapolation and analytic models
employed over the past few years
to construct 3D models of the magnetospheres of accreting pre-main sequence stars.  Each of the methods, analytic,
field extrapolation, and also magnetohydrodynamic which we discuss below, have their own advantages, disadvantages and limitations.  
By writing out the field components for magnetospheres that consist of several high order field components analytically, as in section \ref{analytic},
results that may be missed by numerical models are revealed.  As one example, if we consider the simplified case of
a dipole plus octupole magnetic field, where the dipole and octupole moments are both aligned, then it becomes
clear from equations (\ref{Brdipoct}) and (\ref{Btdipoct}) that a magnetic null point exists where $B_r=B_\theta=0$ at a radius of
\begin{equation}
\frac{r_{null}}{R_\ast} = \left(\frac{3}{4}\frac{B_{oct}}{B_{dip}}\right)^{1/2}.
\label{rnull}
\end{equation}
$r_{null}/R_\ast$ is the radius of circle in a plane tilted by $\beta$ (the tilts of the dipole and octupole moments) towards
the phase that the multipole moments are tilted towards.  Essentially this radius defines the point at where the dipole components
begins to dominate the octupole component as we move away from the stellar surface towards larger radii.  As the polar strength
of the octupole relative to the polar strength of the dipole, $B_{oct}/B_{dip}$, is increased the extent of the region
where the octupole component dominates increases and ``smaller scale'' field lines extending to greater height are found in
the field extrapolations; this can be seen from the plots in Figure \ref{hwplots} for stars with larger $B_{oct}/B_{dip}$ ratios
(see Table \ref{table}).

The major disadvantage of the analytic approach is the inclusion of a large amount of $\ell$ number multipoles rapidly
becomes cumbersome and such models are typically (but not exclusively) limited to axisymmetric fields (e.g. Mohanty $\&$ Shu 2008;
Gregory et al 2010; Gregory 2011).  Furthermore, when the multipole moments are tilted by different amounts relative
to the stellar rotation axis, and/or towards different rotation phases, the field structure (i.e. the paths of the field
lines exterior to the star) must be derived numerically from $B_r/{\rm d}r=B_\theta/(r{\rm d}\theta)=B_\phi/(r\sin{\theta}{\rm d}\phi)$.
Inclusion of non-axisymmetric field modes would further add to the complexity of the analytic work. In such cases 
models of the magnetospheres of stars with complex magnetic fields are best handled numerically.

Numerical field extrapolations from magnetic surface maps have the advantage over the analytic
approach that as many $\ell$ and $m$ modes that can be resolved observationally can be included.  Field extrapolation
is currently the only method where magnetic fields with an observed degree of complexity can be handled and incorporated
into models of the large-scale magnetosphere.  Potential field models are computationally simple to carry out and only require
desktop computing resources.  They also produce unique field topologies (Aly 1987).  However the field topologies obtained
via potential field extrapolation are static.  Time dependent effects, such as the distortion of the large scale magnetic field due to the
interaction with the circumstellar disk, cannot be modeled.  Other non-potential effects, such as flux
emergence or surface differential rotation that has been measured for some accreting T Tauri stars (e.g. Donati et al 2010a, 2011a), 
also cannot be included.

Three-dimensional MHD models of the star-disk interaction with non-dipolar magnetic fields have also been developed by 
Long et al (2007, 2008, 2009, 2011) and Romanova et al (2011).  The main advantage of the MHD approach is the ability to handle the time evolution of both the large-scale
field, and the magnetospheric mass accretion process, that the potential field extrapolation models cannot.
Unfortunately 3D MHD simulations require supercomputer resources; such models are therefore computationally, and fiscally,
expensive.  Likewise, magnetic maps have not yet been incorporated into MHD simulations as grid resolution issues
arise.  A small grid resolution close to the star is required in order to include high order and non-axisymmetric field
components, but at the same time, a large grid able to reach out as far as the inner disk is required.  Instead the
latest generation of 3D MHD models, presented by Romanova et al (2011) and Long et al (2011), begin by
assuming that the star has a potential magnetic field consisting of a tilted dipole plus a tilted octupole component with
field strengths and tilts being adopted from the published magnetic map of the star of interest.  Such dipole-octupole
field structures are then allowed to evolve due to the interaction with the disk.  The
largest scale field lines are torn open and quickly twisted around the rotation axis of the star, see e.g. figure 6 of
Romanova et al (2011), however, the field interior to this that carries gas in columns on to the star retains its initial potential
structure.

MHD models that incorporate the magnetic maps derived from Zeeman-Doppler imaging remain to be carried
out in the future.  Likewise a major new avenue for research is the inclusion of stellar surface differential rotation,
flux emergence and meridional flows into MHD models.  It may turn out that the retention of the potential field
structure found in current simulations is no longer valid once stellar surface transport effects have been accounted for.
Likewise it is unclear whether T Tauri stars undergo magnetic cycles, although there is tentative evidence that
they do and that their large scale magnetospheres evolve over long (possibly several years) timescales (e.g. Donati et al 2011a).

The veracity of the 3D field structures obtained via field extrapolation from the magnetic maps must continue 
to be examined in future.  On going large multi-wavelength programs are useful here (e.g. Argiroffi et al 2011; Donati et al 
2011c).  Based on a field extrapolation model of V2129~Oph Jardine et al (2008), using a simple accretion flow model 
presented in Gregory et al (2007), predicted that a clear signature of accretion related X-ray emission would be detected, which
was subsequently found (Argiroffi et al 2011).  Never-the-less, further tests are underway to determine if the derived field topologies 
genuinely capture the true magnetospheric structure of T Tauri stars.  


\acknowledgements
SGG thanks the organizers for the invitation to give an invited review talk at the meeting, and for financial support.  We
thank Chris Johns-Krull $\&$ Wei Chen for discussions regarding the stellar-disk averaged longitudinal magnetic field
component, and Fred Adams $\&$ Sean Matt for many interesting discussions.  SGG is supported by NASA grant HST-GO-11616.07-A.  
The ``Magnetic Protostars and Planets'' (MaPP) project is supported by the funding agencies of CFHT and TBL (through 
the allocation of telescope time) and by CNRS/INSU in particular, as well as by the French ``Agence Nationale pour la 
Recherche'' (ANR).    

\appendix


\section{Tilted multipole moments in 3D}\label{tiltmaths}

\begin{figure}
\centering
\includegraphics[width=80mm]{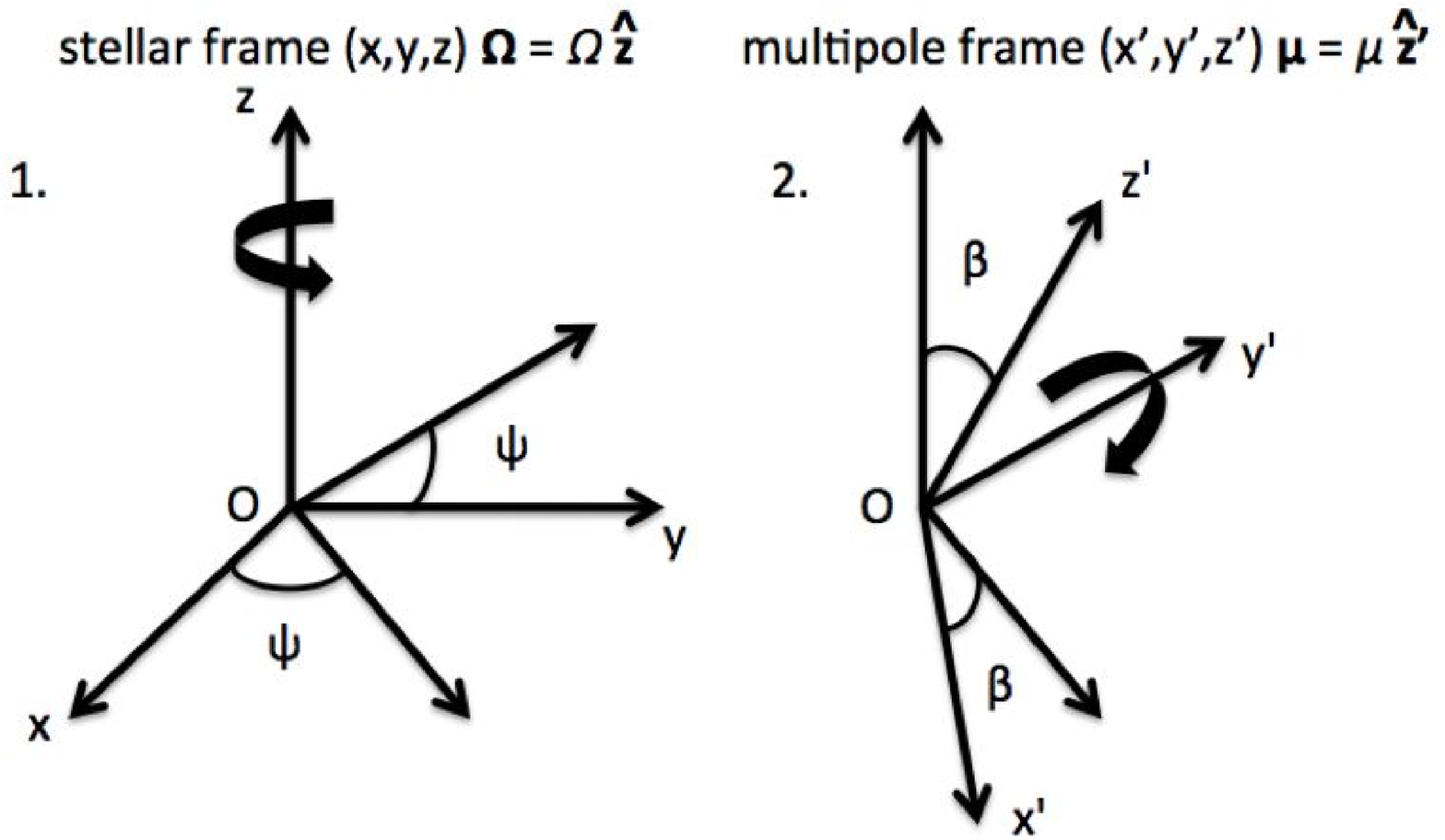}
  \caption{A given multipole moment symmetry axis $\boldsymbol\mu$ is assumed to be tilted by an angle $\beta$, relative to the 
                stellar rotation axis, towards a longitude $\psi$, where $\psi$ is related to the rotation phase via  
                $\psi=(1-{\rm phase}) \times 360^\circ$, assuming the $xz$-plane corresponds to a rotation phase of zero.  The stellar 
                reference frame $(x,y,z)$ is defined such that $\mathbf{\Omega}=\Omega\hat{\mathbf{z}}$,  
                and the multipole frame $(x',y',z')$ such that $\boldsymbol\mu = \mu \hat{\mathbf{z'}}$.  We can transform from the stellar coordinate 
                frame to the multipole frame by 1. rotating the stellar coordinate
                system by $\psi$ around the $z$-axis, followed by 2. a rotation about the new $y$-axis by an angle $\beta$ as illustrated.}
\label{twist}
\end{figure}

We consider two Cartesian coordinate systems (i) the stellar reference frame $(x,y,z)$ where the stellar rotation axis is aligned with the 
$z$-axis ($\mathbf{\Omega}=\Omega\hat{\mathbf{z}}$)
and (ii) the multipole reference frame $(x',y',z')$ where the multipole moment symmetry axis is aligned 
with the $z'$-axis ($\boldsymbol\mu=\mu\hat{\mathbf{z'}}$).  For an arbitrary $\ell$-number multipole Gregory et al. (2010) showed that 
the magnitude of the multipole moment is related to the polar strength of the multipole via,
\begin{equation}
\mu = R_\ast^{\ell+2}B_\ast^{\ell,pole}/(\ell+1).
\end{equation}
We assume that the $xz$-plane corresponds to a rotation phase of zero.  The multipole moment is assumed to be 
tilted by an angle $\beta$, such that $\hat{\mathbf{\Omega}}\cdot\hat{\boldsymbol\mu}=\cos{\beta}$, towards a phase determined by 
another angle $\psi$, the longitude in the stellar reference frame, and where $\hat{\mathbf{\Omega}}$ and $\hat{\boldsymbol\mu}$ are unit vectors 
along $\mathbf{\Omega}$ and $\boldsymbol\mu$ respectively.  
In other words $\beta=45^\circ$ and $\psi=90^\circ$, represents 
a multipole moment tilted by $45^\circ$ from the stellar rotation axis, towards rotation phase 0.75 i.e. $\psi=(1-{\rm phase}) \times 360^\circ$. 
Thus in order to transform from the coordinate system 
that denotes the frame of the star to that which denotes the frame of the multipole we first rotate counterclockwise about 
the $z$-axis (when looking towards the origin) by an angle $\psi$, and then counterclockwise by an angle $\beta$ (when looking 
towards the origin) about the new $y$-axis, as illustrated in Figure \ref{twist} (note that we are rotating the entire coordinate system, 
and not simply a vector within the same fixed coordinate system).  The Cartesian components of the multipole moment
in the stellar reference frame, where $\boldsymbol\mu=\mu_x\hat{\mathbf{x}}+ \mu_y\hat{\mathbf{y}} + \mu_z\hat{\mathbf{z}}$, are 
{\setlength{\mathindent}{-10pt}
\begin{eqnarray}
&&\left(
\begin{array}{c}
\mu_x \\
\mu_y \\
\mu_z
\end{array}
\right) = \nonumber \\
&& \left(
\begin{array}{ccc}
\cos{\psi} & -\sin{\psi} & 0 \\
\sin{\psi} & \cos{\psi} & 0 \\
0 & 0 & 1
\end{array} 
\right)\left(
\begin{array}{ccc}
\cos{\beta} & 0 & \sin{\beta} \\
0 & 1 & 0 \\
-\sin{\beta} & 0 & \cos{\beta}
\end{array}
\right)\left(
\begin{array}{c}
0 \\
0 \\
\mu \\
\end{array}
\right).
\end{eqnarray}}
The vector components of $\boldsymbol\mu$ can then be converted from the Cartesian to a spherical coordinate system $(r,\theta,\phi)$ where 
$r=0$ corresponds to the center of the star, $\phi=0$ corresponds to the stellar $xz$-plane and $\theta$ is the co-latitude in 
the stellar reference frame. 
Noting that $\boldsymbol\mu=\mu_x\hat{\mathbf{x}}+ \mu_y\hat{\mathbf{y}} + \mu_z\hat{\mathbf{z}}=\mu_r\hat{\mathbf{r}}+
\mu_\theta\hat{\boldsymbol\theta}+\mu_\phi\hat{\boldsymbol\phi}$, then
{\setlength{\mathindent}{0pt}
\begin{eqnarray}
\left(
\begin{array}{c}
\mu_r \\
\mu_\theta \\
\mu_\phi
\end{array}
\right) &=&
\left(
\begin{array}{ccc}
\sin{\theta}\cos{\phi} & \sin{\theta}\sin{\phi} & \cos{\theta} \\
\cos{\theta}\cos{\phi} & \cos{\theta}\sin{\phi} & -\sin{\theta} \\
-\sin{\phi} & \cos{\phi} & 0
\end{array}
\right)\left(
\begin{array}{c}
\mu_x \\
\mu_y \\
\mu_z
\end{array}
\right).
\end{eqnarray}}
The spherical components of the tilted multipole moment in the stellar reference frame are thus given by 
equations (\ref{mu_r}-\ref{mu_phi}).


\section{Stellar disk averaged longitudinal magnetic field component}\label{long}
Goossens (1979) derived an expression for the stellar disk averaged longitudinal magnetic field component $B_{lon}$ arising from a 
multipolar stellar magnetic field, see his equation (20) with the `W' terms for a linear limb darkening law from his 
equations (25a-d) - see also Stibbs (1950) and Schwarzschild (1950).  It is applicable to the case where all of the 
multipole moments are aligned with each other, and which are viewed at an angle 
$\alpha$ to the magnetic axis.  The first term in Goossens' equation (20) is the dipole term, the second term is the quadrupole term, the 
third term represents the contribution from the higher order even $\ell$ number multipoles
($\ell=$ 4 [hexadecapole], 6, 8 $\ldots$), and the fourth term is the contribution from the higher order odd $\ell$ number multipoles 
($\ell=$ 3 [octupole], 5 [dotriacontapole], 7 $\ldots$).  Goossens then goes on to derive the stellar disk averaged longitudinal 
field component for a dipole-quadrupole magnetic field, but this can be extended to any arbitrary combination of high order $\ell$ number
multipoles.

By comparing the magnetostatic potential in Goossens (1979) [his equation (4)] with that used in Gregory et al. (2010) 
[their equations (3.18) and (3.27)], the `A' terms in Goossens (1979) equation (20) are 
related to the polar field strengths of the particular multipoles via,
\begin{equation}
A_n = M_\ell = \frac{1}{(\ell+1)}R_\ast^{\ell+2}B_{\ast}^{\ell,{\rm pole}}, \nonumber
\end{equation}
where $B_\ast^{\ell,{\rm pole}}$ is the polar field strength of the $\ell$th multipole, and we rewrite $B_\ast^{1,{\rm pole}}\equiv B_{dip}$, 
$B_\ast^{2,{\rm pole}}\equiv B_{quad}$, and $B_\ast^{3,{\rm pole}}\equiv B_{oct}$ for brevity.  
Thus, the `A' terms in Goossens equation (20) can be re-written as 
\begin{eqnarray}
A_1 &=& \frac{1}{2}R_\ast^3B_{dip}  \nonumber \\
A_2 &=& \frac{1}{3}R_\ast^4B_{quad}  \nonumber \\
A_{2k} &=& \frac{1}{(2k+1)}R_\ast^{2k+2}B_\ast^{2k,{\rm pole}} \nonumber \\
A_{2k+1} &=& \frac{1}{(2k+2)}R_\ast^{2k+3}B_\ast^{2k+1,{\rm pole}} \nonumber
\end{eqnarray}
where the first term is the dipole term, the second the quadrupole term, the third term the higher order even $\ell$ number multipoles, and the 
fourth term the higher order odd $\ell$ number multipoles.  Equation (20) of Goossens can then be re-expressed by replacing the `A' and `W' terms,
\begin{eqnarray}
B_{lon} &=& \frac{1}{20}B_{dip}\frac{(15+u)}{(3-u)}P_1(\cos{\alpha}) \nonumber \\
&& +\frac{1}{4}B_{quad}\frac{u}{(3-u)}P_2(\cos{\alpha}) \nonumber \\
&& +\frac{3u}{(3-u)} \sum_{k=2}^{\infty}C_k \frac{B_\ast^{2k,{\rm pole}}}{(2k+1)}P_{2k}(\cos{\alpha}) \nonumber \\
&& +\frac{3(1-u)}{(3-u)}\sum_{k=1}^{\infty}D_k\frac{B_\ast^{2k+1,{\rm pole}}}{(2k+2)}P_{2k+1}(\cos{\alpha}),
\label{goo}
\end{eqnarray}
where $P_j(\cos{\alpha})$ are the Legendre polynomials, $\alpha$ is the angle between the magnetic axis (here it is assumed all the 
multipole moments are aligned with each other) and the observer's line-of-sight, and
\begin{eqnarray}
C_k &=& (-1)^{k+1} \frac{(2k+1)(2k-3)!!}{2^k(k+2)!}  \nonumber \\
D_k &=& (-1)^k\frac{(k+1)(2k-1)!!}{2^k(k+2)!} \nonumber
\end{eqnarray}
with $(2k-1)!!=(2k-1)(2k-3)(2k-5)\ldots 5\times 3 \times 1$. 

\subsection{Dipole-octupole longitudinal magnetic field component}
The dipole longitudinal field component is the first term of equation (\ref{goo}), and derivations can be found in Schwarzschild (1950) and 
Stibbs (1950). The octupole magnetic field term is the $k=1$ term in the final summation term of equation (\ref{goo}).  Thus the disk 
averaged longitudinal magnetic field component for a pure octupole is,
\begin{eqnarray}
B^{oct}_{lon} &=& \frac{1}{8}B_{oct}\frac{(u-1)}{(3-u)}P_3(\cos{\alpha}) \nonumber \\
                     &=& \frac{1}{16}B_{oct}\frac{(u-1)}{(3-u)}(5\cos^3{\alpha}-3\cos{\alpha}) \nonumber
\end{eqnarray}
where $\alpha$ is the angle between the octupole moment and the observer's line-of-sight.  $\alpha$ is related to the stellar inclination $i$ and 
the angle between the stellar rotation pole and the multipole moment $\beta$ via,
\begin{equation}
\cos{\alpha}=\cos{i}\cos{\beta}+\sin{i}\sin{\beta}\cos{\phi}
\label{sph}
\end{equation}
where $\phi$ is the azimuthal angle (the longitude) of the magnetic axis at some time $t$ i.e. $\phi=2\pi(1-\Phi)$ where $\Phi$ is the 
rotation phase (see the discussion in Goossens (1979) around his equations (23) and (24)). Equation (\ref{sph}) can be derived 
by considering the circle swept out by the magnetic axis rotating about the stellar
rotation axis (it is just the spherical cosine law).  If at time $t=0$, i.e. rotation phase $\Phi=0$, the magnetic axis is tilted towards some
rotation phase $\Phi_0$ then $\cos{\phi}=\cos{(2\pi[1-\Phi]+2\pi[1-\Phi_0])}=\cos{(2\pi[\Phi-\Phi_0])}$, thus,
\begin{equation}
\cos{\alpha}=\cos{i}\cos{\beta}+\sin{i}\sin{\beta}\cos{(2\pi[\Phi-\Phi_0])}.
\end{equation}
The stellar disk averaged longitudinal magnetic field component for an octupole is then,
{\setlength{\mathindent}{0pt}
\begin{eqnarray}
&& B_{lon}^{oct}(\Phi) = \frac{1}{16}B_{oct} \left(\frac{u-1}{3-u}\right)\times \nonumber \\
&& \Big\{5(\cos{\beta_{oct}}\cos{i}+\sin{\beta_{oct}}\sin{i}\cos{[2\pi(\Phi-\Phi_{oct})]})^3 \nonumber \\   
&&-3(\cos{\beta_{oct}}\cos{i}+\sin{\beta_{oct}}\sin{i}\cos{[2\pi(\Phi-\Phi_{oct})]})\Big\}.
\end{eqnarray}}

Goossens (1979) consider the dipole-quadrupole case where the dipole and quadrupole moments are tilted by different 
amounts towards different rotation phases (there is a $\sin{i}$ term missing in the second term of his equation (35)).  The Goossens
work is straightforward to extend to dipole-octupole magnetic fields, with the dipole and octupule tilted towards different rotation
phases and by different amounts relative to the stellar rotation axis.  Taking the dipole and octupole components of equation (\ref{goo})
\begin{eqnarray}
B_{lon}(\Phi)&=&\frac{1}{20}B_{dip}\frac{(15+u)}{(3-u)}\cos{\alpha_{dip}} \nonumber \\
&+&\frac{1}{16}B_{oct}\frac{(u-1)}{(3-u)}(5\cos^3{\alpha_{oct}}-3\cos{\alpha_{oct}})
\end{eqnarray}
where $\alpha_{dip}$ and $\alpha_{oct}$ are the angles between the dipole moment and the stellar rotation axis, and the octupole 
moment and the stellar rotation axis, respectively.  In the notation adopted by Goossens, see their equation (35), $\lambda$ is the 
angle between the plane that contains the dipole moment and the stellar rotation axis, and the plane that contains the quadrupole 
moment (octupole moment in our case) and the stellar rotation axis.  Thus, in the Goossens notation, at time $t=0$ if the dipole moment 
is titled towards longitude 
$\phi_0$ then the octupole (or quadrupole in the Goossens paper) moment is tilted towards longitude $\phi_0+\lambda$; in other words we can write,
{\setlength{\mathindent}{0pt}
\begin{eqnarray}
&&\cos{\alpha_{dip}} = \nonumber \\
&& \cos{i}\cos{\beta_{dip}}+\sin{i}\sin{\beta_{dip}}\cos{(2\pi[\Phi-\Phi_{dip}])} \\
&&\cos{\alpha_{oct}} = \nonumber \\
&& \cos{i}\cos{\beta_{oct}}+\sin{i}\sin{\beta_{oct}}\cos{(2\pi[\Phi-\Phi_{oct}])}.
\end{eqnarray}} 
Putting the equations in this section together gives equation (\ref{blondipoct}), the general result for the dipole-octupole stellar disk averaged 
longitudinal magnetic field component as a function of rotation phase.


\end{document}